\documentclass[journal]{IEEEtran}
\usepackage[T1]{fontenc}
\usepackage[latin9]{inputenc}
\usepackage{array}
\usepackage{mathrsfs}
\usepackage{multirow}
\usepackage{amsmath}
\usepackage{amsthm}
\usepackage{amssymb}
\usepackage{graphicx}
\usepackage[unicode=true,
 bookmarks=true,bookmarksnumbered=true,bookmarksopen=true,bookmarksopenlevel=1,
 breaklinks=false,pdfborder={0 0 0},pdfborderstyle={},backref=false,colorlinks=false]
 {hyperref}
\hypersetup{pdftitle={Your Title},
 pdfauthor={Your Name},
 pdfpagelayout=OneColumn, pdfnewwindow=true, pdfstartview=XYZ, plainpages=false}

\makeatletter

\providecommand{\tabularnewline}{\\}

\theoremstyle{plain}
\newtheorem{thm}{\protect\theoremname}
\theoremstyle{remark}
\newtheorem{rem}[thm]{\protect\remarkname}

\usepackage[caption=false,font=footnotesize]{subfig}

\makeatother

\providecommand{\remarkname}{Remark}
\providecommand{\theoremname}{Theorem}

\begin{document}
\title{Modeling and Architecture Design of Reconfigurable Intelligent Surfaces
Using \\
Scattering Parameter Network Analysis}
\author{Shanpu Shen,~\IEEEmembership{Member,~IEEE}, Bruno Clerckx,~\IEEEmembership{Senior Member,~IEEE},
and Ross Murch,~\IEEEmembership{Fellow,~IEEE}\thanks{Manuscript received; This work was supported by the Hong Kong Research
Grants Council with the Collaborative Research Fund grant C6012-20G.
\textit{(Corresponding author: Shanpu Shen.)}}\thanks{S. Shen is with the Department of Electronic and Computer Engineering,
The Hong Kong University of Science and Technology, Clear Water Bay,
Kowloon, Hong Kong (e-mail: sshenaa@connect.ust.hk).}\thanks{B. Clerckx is with the Department of Electrical and Electronic Engineering,
Imperial College London, London SW7 2AZ, U.K. (e-mail: b.clerckx@imperial.ac.uk).}\thanks{R. Murch is with the Department of Electronic and Computer Engineering
and the Institute of Advanced Study, The Hong Kong University of Science
and Technology, Clear Water Bay, Kowloon, Hong Kong (e-mail: eermurch@ust.hk).}}
\maketitle
\begin{abstract}
Reconfigurable intelligent surfaces (RISs) are an emerging technology
for future wireless communication. The vast majority of recent research
on RIS has focused on system level optimizations. However, developing
straightforward and tractable electromagnetic models that are suitable
for RIS aided communication modeling remains an open issue. In this
paper, we address this issue and derive communication models by using
rigorous scattering parameter network analysis. We also propose new
RIS architectures based on group and fully connected reconfigurable
impedance networks that can adjust not only the phases but also the
magnitudes of the impinging waves, which are more general and more
efficient than conventional single connected reconfigurable impedance
network that only adjusts the phases of the impinging waves. In addition,
the scaling law of the received signal power of an RIS aided system
with reconfigurable impedance networks is also derived. Compared with
the single connected reconfigurable impedance network, our group and
fully connected reconfigurable impedance network can increase the
received signal power by up to 62\%, or maintain the same received
signal power with a number of RIS elements reduced by up to 21\%.
We also investigate the proposed architecture in deployments with
distance-dependent pathloss and Rician fading channel, and show that
the proposed group and fully connected reconfigurable impedance networks
outperform the single connected case by up to 34\% and 48\%, respectively.
\end{abstract}


\section{Introduction}

\IEEEPARstart{R}{econfigurable} intelligent surfaces (RISs), also
known as intelligent reflecting surfaces, have gained popularity as
a revolutionary technology for achieving spectrum efficient, energy
efficient, and cost effective wireless communication \cite{2019_IEEEAccess_RIS_review},
\cite{2019_ArXiv_IRS_Toward_YCL}, \cite{2020_ArXiv_IRS_Tutorial_QWu}.
RISs can alter or reconfigure the propagation environment so that
the performance of wireless communications can be significantly improved
\cite{2020_CM_TowardIRS_QWu}. For these reasons RIS is considered
a potentially important technology for use in future 6G communications
\cite{rajatheva2020white}.

An RIS consists of a large number of reconfigurable passive elements,
where each element is able to introduce phase shift to the scattered
signal. By collaboratively adjusting the phase shifts of all passive
elements of the RIS, the scattered signals can add coherently with
the signals from other paths at the desired receiver to boost the
received signal power. Alternatively the signal paths can also be
made to destructively add at non-intended receivers to suppress interference
as well as enhance security and privacy. In contrast with amplify-and-forward
(AF) relay technology \cite{2010_TIT_AF}, RIS has several advantages
including low cost, low power consumption, contributing no active
additive thermal noise or self-interference enabling full-duplexing
operation. In addition, RIS exhibits potential features such as being
low profile, light weight, and having conformal geometry, making them
straightforward to deploy.

Due to the potential advantages, RIS has been investigated in various
wireless communication systems including multiple-input single-output
(MISO) \cite{2019_TWC_RIS_CHuang}, \cite{2019_TWC_IRS_QWu}, multiple-input
multiple-output (MIMO) \cite{2020_WCL_IRS_MIMO_Beamform}, multicell
\cite{2020_TWC_IRS_MIMO_Multicell}, and multigroup multicast \cite{2020_TSP_IRS_Multigroup_Multicast}.
Specifically, for RIS aided MISO systems, in \cite{2019_TWC_RIS_CHuang}
energy efficiency is maximized by optimizing phase shifts while in
\cite{2019_TWC_IRS_QWu} the transmit power is minimized by jointly
optimizing active and passive beamforming. For RIS aided MIMO systems,
in \cite{2020_WCL_IRS_MIMO_Beamform} spectral efficiency is maximized
based on a sum-path-gain maximization criterion. For RIS aided multicell
MIMO systems, the weighted sum rate maximization problem is considered
in \cite{2020_TWC_IRS_MIMO_Multicell} and it is shown that the cell-edge
performance can be significantly enhanced by RIS. For RIS aided multigroup
multicast communication systems, the sum rate of multiple multicasting
groups is maximized in \cite{2020_TSP_IRS_Multigroup_Multicast}.
Furthermore, RIS has also been investigated to provide performance
enhancement in orthogonal frequency division multiplexing (OFDM) \cite{2020_ToC_IRS_OFDM},
\cite{2020_WCL_IRS_OFDM}, non-orthogonal multiple access (NOMA) \cite{2020_CL_IRS_NOMA},
and emerging areas such as secure wireless communication \cite{2020_JSAC_IRS_Robust_Secure},
backscatter communication \cite{2020_WCL_LIS_BackScatter}, simultaneous
wireless information and power transfer (SWIPT) \cite{2020_JSCA_IRS_SWIPT},
\cite{2020_WCL_IRS_SWIPT_QWu}, spectrum sharing \cite{2020_CL_IRS_spectrum_Sharing},
cognitive radio \cite{2021_ToC_IRS_CognitiveRadio}, unmanned aerial
vehicle (UAV) communication \cite{2020_IEEEAccess_IRS_UAV}, millimeter
wave \cite{2020_WCL_IRS_mmwave_Secure}, \cite{2020_WCL_IRS_mmwave_Terahertz},
and mobile edge computing \cite{2020_JSAC_IRS_MEC}. In addition,
research on optimizing RIS aided wireless systems with discrete phase
shifts \cite{2019_PACRIM_IRS_DiscretePhaseShift}, \cite{2020_ToC_IRS_Discrete_Phase},
statistic and imperfect channel state information (CSI) \cite{2019_TVT_LIS_Statistic_CSI},
\cite{2020_TSP_IRS_GuiZhou} and deep learning \cite{2020_JSAC_RIS_DeepReinLearning},
\cite{2020_WCL_IRS_DRL} has also been conducted.

While the vast majority of research on RIS has been devoted to system
level optimization \cite{2019_TWC_RIS_CHuang}-\cite{2020_WCL_IRS_DRL},
developing models that satisfy the necessary electromagnetic (EM)
equations while providing tractable and useful RIS aided communication
models still remains an open problem. There are only a few published
results analyzing the physical and EM properties of RIS. In \cite{2020_WCL_IRS_Physical_Prop_Larsson},
physical optics is utilized to obtain expressions for the scattered
field from a passive metallic surface and accordingly an RIS pathloss
model is derived. In addition, in \cite{2020_TWC_RIS_PathLossModelMeasure}
a free-space pathloss model for an RIS aided wireless communication
is introduced from the perspective of EM theory and experimentally
verified in a microwave anechoic chamber. In addition to the pathloss
models, practical phase shift models of RIS accounting for lumped
inductance and capacitance are proposed in \cite{2020_ICC_IRS_PhaseShiftModel},
\cite{2020_CL_IRS_PracticalModel}, and the optimization based on
the practical phase shift models are also provided. However, the limitation
of \cite{2020_WCL_IRS_Physical_Prop_Larsson}, \cite{2020_TWC_RIS_PathLossModelMeasure},
\cite{2020_ICC_IRS_PhaseShiftModel}, \cite{2020_CL_IRS_PracticalModel}
is that they only focus on very specific physical and EM properties
of RIS. Therefore, how to derive a straightforward and tractable yet
EM based RIS aided communication model remains an open problem.

In addition to the RIS aided communication modeling issue, another
challenge in enabling the promise of RIS is that the signal power
received from the RIS is limited (or equivalently the composite transmitter-RIS-receiver
channel gain is very low). As shown in \cite{2020_WCL_IRS_Physical_Prop_Larsson},
the signal power received from an RIS is proportional to the square
of RIS area and to $1/\left(d_{i}r\right)^{2}$ where $d_{i}$ is
the distance between the transmitter and RIS and $r$ is the distance
between the RIS and receiver. In addition, comparisons with massive
MIMO \cite{2019_CAMSAP_IRS_PowerScaling} and decode-and-forward relays
\cite{2020_WCL_IRS_DF} indicate that RIS needs a large number of
elements to be competitive. Therefore, it remains a challenge to develop
an efficient RIS architecture to improve the received signal power.

In this paper, we derive a straightforward and tractable yet EM based
RIS aided communication model using a rigorous scattering parameter
network analysis. It has been inspired by previous results on MIMO
antennas \cite{2004_TWC_Jensen} and here it is extended to RIS. We
also propose efficient RIS architectures, namely fully connected and
group connected reconfigurable impedance networks, to improve the
received signal power. The contributions of the paper are summarized
as follows.

\textit{First}, we derive a physical and EM compliant RIS aided communication
model using scattering parameter network analysis. This is the first
paper to characterize and model RIS from the perspective of scattering
parameters. Using scattering parameters is beneficial for accounting
for the scattering mechanism of RIS. The derived model is general
enough that it accounts for the impedance mismatch and mutual coupling
at the transmitter, RIS, and receiver. Additionally, assuming perfect
matching and no mutual coupling, we can simplify the model and achieve
a straightforward and tractable RIS aided communication model. The
conventional RIS model used in \cite{2019_TWC_RIS_CHuang}-\cite{2020_WCL_IRS_DRL}
for example is a particular instance of the proposed model.

\textit{Second}, we investigate the RIS architecture and propose two
new architectures based on fully connected and group connected reconfigurable
impedance networks, that are respectively modeled using complex symmetric
unitary and block diagonal matrices with each block being complex
symmetric unitary. Those two architectures are more general than the
conventional single connected reconfigurable impedance network used
in \cite{2019_TWC_RIS_CHuang}-\cite{2020_WCL_IRS_DRL}, which is
modeled using a diagonal matrix with each entry having a unit modulus.
In sharp contrast with the conventional single connected architecture
that only adjusts the phases of the impinging waves, our proposed
fully and group connected architectures can adjust not only the phases
but also the magnitudes of the impinging waves. This is the first
paper to introduce fully connected and group connected networks and
show the benefit over a single connected network.

\textit{Third}, we derive the scaling law of the received signal power
of a single-input single-output (SISO) RIS aided system as a function
of the number of RIS elements. Both line-of-sight (LoS) and Rayleigh
fading channels have been considered. It shows the power gain of the
fully connected and group connected reconfigurable impedance network
over the single connected reconfigurable impedance network in Rayleigh
fading channels can be up to 1.62. Given the same received signal
power, it is shown that using fully connected and group connected
reconfigurable impedance networks can reduce the number of RIS elements
by up to 21\%, which is beneficial for reducing the cost and area
of RIS, especially when the number of RIS elements is large. In addition,
it shows that the group connected reconfigurable impedance network
with small group size can provide most of the performance enhancement
and come close to the fully connected case while maintaining low complexity.

\textit{Fourth}, we optimize the scattering matrix of the reconfigurable
impedance network in RIS to maximize the received signal power in
the SISO RIS aided system. We also evaluate the received signal power
in channel models with distance-dependent pathloss and Rician fading
channels, which is more general than the channel model used in previous
scaling law analysis. The numerical results show that the fully connected
and group connected reconfigurable impedance networks can increase
the received signal power by up to 48\% and 34\%, respectively.

\textit{Organization}: Section II provides the scattering parameter
network analysis and proposes the fully connected and group connected
reconfigurable impedance network. Section III provides the RIS aided
communication model. Section IV provides the scaling laws for the
SISO RIS aided system. Section V evaluates the performance of the
proposed RIS and Section VI provides conclusions and details possible
future work.

\textit{Notation}: Bold lower and upper letters denote vectors and
matrices, respectively. Letters not in bold font represent scalars.
$\Re\left\{ a\right\} $, $\left|a\right|$, and $\arg\left(a\right)$
refer to the real part, modulus, and phase of a complex scalar $a$,
respectively. $\left[\mathrm{\mathbf{a}}\right]_{i}$ and $\left\Vert \mathbf{a}\right\Vert $
refer to the $i$th element and $l_{2}-$norm of vector $\mathrm{\mathbf{a}}$,
respectively. $\mathrm{\mathbf{A}}^{T}$, $\mathrm{\mathbf{A}}^{H}$,
and $\left[\mathrm{\mathbf{A}}\right]_{i,j}$ refer to the transpose,
conjugate transpose, and $\left(i,j\right)$th element a matrix $\mathrm{\mathbf{A}}$,
respectively. $\mathbb{R}$ and $\mathbb{C}$ denote real and complex
number set, respectively. $j=\sqrt{-1}$ denotes imaginary unit. $\mathbf{0}$
and $\mathbf{I}$ denote an all-zero matrix and an identity matrix,
respectively, with appropriate dimensions. $\chi_{k}$ denotes the
chi distribution with $k$ degrees of freedom. $\mathcal{CN}\left(\mathbf{0},\mathbf{I}\right)$
denotes the distribution of a circularly symmetric complex Gaussian
random vector with mean vector $\mathbf{0}$ and covariance matrix
$\mathbf{I}$ and $\sim$ stands for \textquotedblleft distributed
as\textquotedblright . $\mathbf{A}\preceq\mathbf{B}$ means that $\mathbf{B}-\mathbf{A}$
is positive semi-definite. diag$\left(a_{1},...,a_{N}\right)$ refers
to a diagonal matrix with diagonal elements being $a_{1},...,a_{N}$.
diag$\left(\mathrm{\mathbf{A}}_{1},...,\mathrm{\mathbf{A}}_{N}\right)$
refers to a block diagonal matrix with blocks being $\mathrm{\mathbf{A}}_{1},...,\mathrm{\mathbf{A}}_{N}$.

\section{Network Analysis}

In this section, we use scattering parameter network theory to analyze
RIS aided wireless communication systems. A brief review of the basic
concepts of reflection coefficient and scattering parameter network
theory is provided in the Appendix.

\begin{figure}[t]
\begin{centering}
\includegraphics[width=8.5cm]{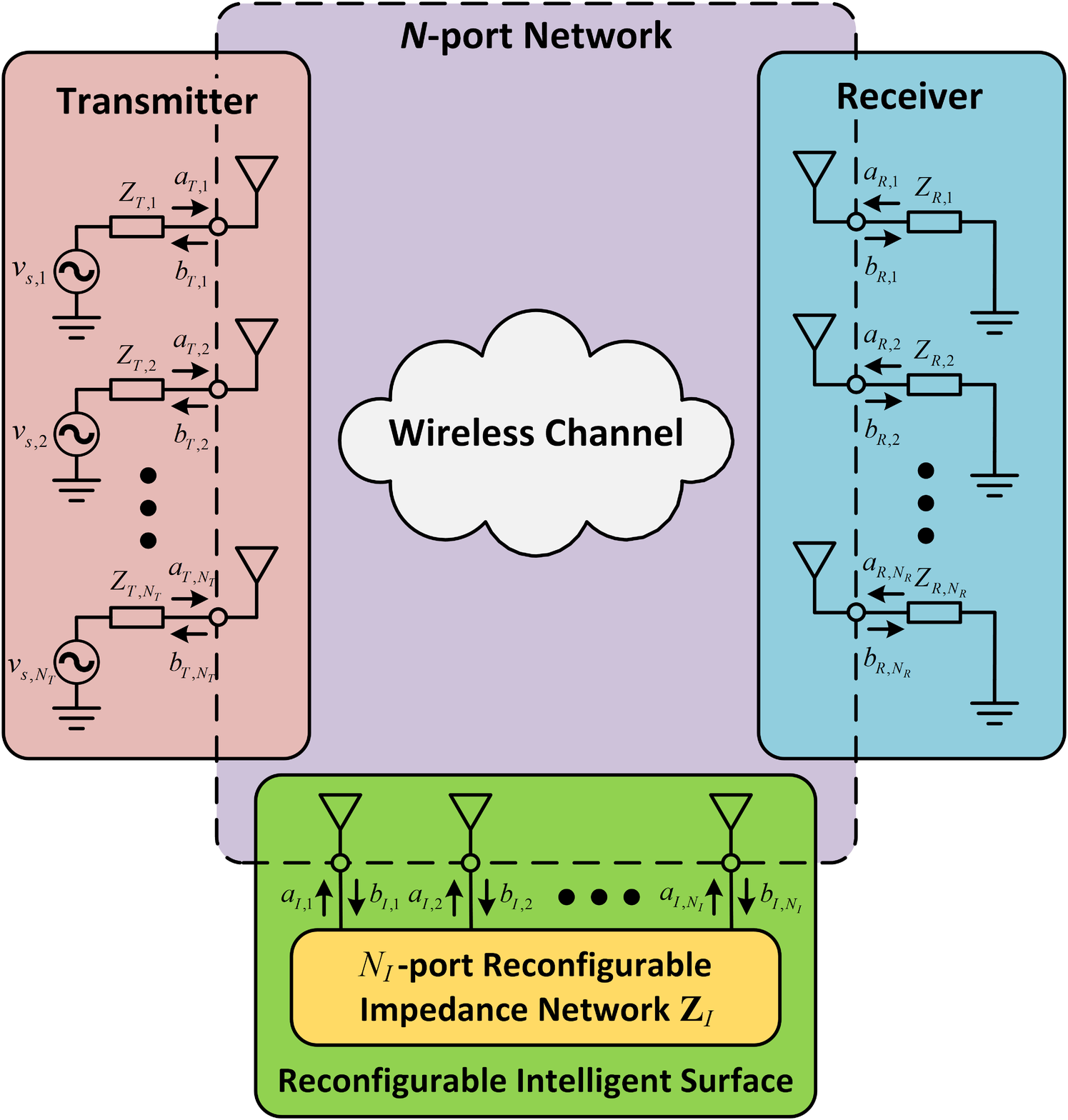}
\par\end{centering}
\caption{\label{fig:System-diagram}System diagram of RIS aided wireless communication
system.}
\end{figure}

A system diagram of an RIS aided wireless communication system is
shown in Fig. \ref{fig:System-diagram}. There are $N_{T}$ antennas
at the transmitter, $N_{R}$ antennas at the receiver, and $N_{I}$
antennas at the RIS. In total, there are $N=N_{T}+N_{I}+N_{R}$ antennas
embedded in the wireless channel, which can be modeled as an $N$-port
network. The $N$-port network can be characterized by a scattering
matrix $\mathbf{S}$ where $\mathbf{S}\in\mathbb{C}^{N\times N}$
and $\mathbf{S}=\mathbf{S}^{T}$ due to the reciprocity, so that we
have 
\begin{equation}
\mathbf{b}=\mathbf{S}\mathbf{a},\label{eq:b=00003DSa}
\end{equation}
where $\mathbf{a},\mathbf{b}\in\mathbb{C}^{N\times1}$ refer to the
incident and reflected waves at the $N$ ports, respectively. Particularly,
we can partition $\mathbf{a}$ and $\mathbf{b}$ as 
\begin{equation}
\mathbf{a}=\left[\begin{array}{l}
\mathbf{a}_{T}\\
\mathbf{a}_{I}\\
\mathbf{a}_{R}
\end{array}\right],\:\mathbf{b}=\left[\begin{array}{l}
\mathbf{b}_{T}\\
\mathbf{b}_{I}\\
\mathbf{b}_{R}
\end{array}\right],
\end{equation}
where the subscripts $T$, $I$, and $R$ refer to the transmitter,
RIS, and receiver, respectively, and $\mathbf{a}_{i}=\left[a_{i,1},a_{i,2},\ldots,a_{i,N_{i}}\right]^{T}\in\mathbb{C}^{N_{i}\times1}$
and $\mathbf{b}_{i}=\left[b_{i,1},b_{i,2},\ldots,b_{i,N_{i}}\right]^{T}\in\mathbb{C}^{N_{i}\times1}$
for $i\in\left\{ T,I,R\right\} $ refer to the incident and reflected
waves of the antennas at the transmitter/RIS/receiver. Accordingly,
we can partition $\mathbf{S}$ as 
\begin{equation}
\mathbf{S}=\left[\begin{array}{ccc}
\mathbf{S}_{TT} & \mathbf{S}_{TI} & \mathbf{S}_{TR}\\
\mathbf{S}_{IT} & \mathbf{S}_{II} & \mathbf{S}_{IR}\\
\mathbf{S}_{RT} & \mathbf{S}_{RI} & \mathbf{S}_{RR}
\end{array}\right],
\end{equation}
where $\mathbf{S}_{TT}\in\mathbb{C}^{N_{T}\times N_{T}}$, $\mathbf{S}_{II}\in\mathbb{C}^{N_{I}\times N_{I}}$,
$\mathbf{S}_{RR}\in\mathbb{C}^{N_{R}\times N_{R}}$ refer to the scattering
matrices of the antenna arrays at the transmitter, RIS, and receiver,
respectively. The diagonal entries of $\mathbf{S}_{TT}$, $\mathbf{S}_{II}$,
and $\mathbf{S}_{RR}$ refer to the antenna reflection coefficients
while the off-diagonal entries refer to antenna mutual coupling. $\mathbf{S}_{RT}\in\mathbb{C}^{N_{R}\times N_{T}}$,
$\mathbf{S}_{IT}\in\mathbb{C}^{N_{I}\times N_{T}}$, and $\mathbf{S}_{RI}\in\mathbb{C}^{N_{R}\times N_{I}}$
refer to the transmission scattering matrices from the transmitter
to receiver, from the transmitter to RIS, and from RIS to the receiver,
respectively.

\subsection{Transmitter and Receiver}

At the transmitter, for $n_{T}=1,\ldots,N_{T}$, the $n_{T}$th transmit
antenna is connected in series with a voltage source, denoted as $v_{s,n_{T}}$,
and a source impedance, denoted as $Z_{T,n_{T}}$. Therefore, $\mathbf{a}_{T}$
and $\mathbf{b}_{T}$ are related by 
\begin{align}
\mathbf{a}_{T} & =\mathbf{b}_{s,T}+\mathbf{\Gamma}_{T}\mathbf{b}_{T},\label{eq:atbt}
\end{align}
where $\mathbf{b}_{s,T}=\frac{1}{2}\left[v_{s,1},v_{s,2},\ldots,v_{s,N_{T}}\right]^{T}\in\mathbb{C}^{N_{T}\times1}$
refers to the wave source vector and $\mathbf{\Gamma}_{T}\in\mathbb{C}^{N_{T}\times N_{T}}$
is a diagonal matrix with its $\left(n_{T},n_{T}\right)$th entry
referring to the reflection coefficient of the $n_{T}$th source impedance,
i.e.
\begin{equation}
\left[\mathbf{\Gamma}_{T}\right]_{n_{T},n_{T}}=\frac{Z_{T,n_{T}}-Z_{0}}{Z_{T,n_{T}}+Z_{0}},
\end{equation}
where $Z_{0}$ refers to the reference impedance used for computing
the scattering parameter, and usually set as $Z_{0}=50\:\Omega$.

At the receiver, for $n_{R}=1,\ldots,N_{R}$, the $n_{R}$th receive
antenna is connected in series with a load impedance, denoted as $Z_{R,n_{R}}$.
Therefore, $\mathbf{a}_{R}$ and $\mathbf{b}_{R}$ are related by
\begin{align}
\mathbf{a}_{R} & =\mathbf{\Gamma}_{R}\mathbf{b}_{R},\label{eq:arbr}
\end{align}
where $\mathbf{\Gamma}_{R}\in\mathbb{C}^{N_{R}\times N_{R}}$ is a
diagonal matrix with its $\left(n_{R},n_{R}\right)$th entry referring
to the reflection coefficient of the $n_{R}$th load impedance, i.e.
\begin{equation}
\left[\mathbf{\Gamma}_{R}\right]_{n_{R},n_{R}}=\frac{Z_{R,n_{R}}-Z_{0}}{Z_{R,n_{R}}+Z_{0}}.
\end{equation}

\subsection{Reconfigurable Intelligent Surface}

At the RIS, the $N_{I}$ antennas are connected to a $N_{I}$-port
reconfigurable impedance network. Therefore, $\mathbf{a}_{I}$ and
$\mathbf{b}_{I}$ are related by
\begin{align}
\mathbf{a}_{I} & =\boldsymbol{\Theta}\mathbf{b}_{I},\label{eq:aibi}
\end{align}
where $\boldsymbol{\Theta}\in\mathbb{C}^{N_{I}\times N_{I}}$ refers
to the scattering matrix of the $N_{I}$-port reconfigurable impedance
network. According to \cite{pozar2009microwave}, $\boldsymbol{\Theta}$
can be expressed as
\begin{equation}
\boldsymbol{\Theta}=\left(\mathbf{Z}_{I}+Z_{0}\mathbf{I}\right)^{-1}\left(\mathbf{Z}_{I}-Z_{0}\mathbf{I}\right),\label{eq:gammaI =00003D (ZI-Z0)(ZI+Z0)-1}
\end{equation}
where $\mathbf{Z}_{I}\in\mathbb{C}^{N_{I}\times N_{I}}$ refers to
the impedance matrix of the $N_{I}$-port reconfigurable impedance
network. The $N_{I}$-port reconfigurable impedance network is constructed
with reconfigurable and passive elements so that it can reflect the
incident signal with a reconfiguration that can be adapted to the
channel. The $N_{I}$-port reconfigurable impedance network is also
reciprocal so that we have symmetry where $\mathbf{Z}_{I}=\mathbf{Z}_{I}^{T}$
and $\boldsymbol{\Theta}=\boldsymbol{\Theta}^{T}$. According to the
circuit network topology, the $N_{I}$-port reconfigurable impedance
network can be classified into three categories.

\begin{figure}[t]
\begin{centering}
\includegraphics[width=8.5cm]{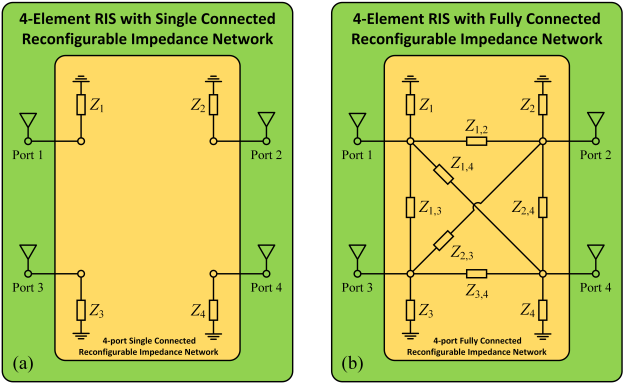}
\par\end{centering}
\caption{\label{fig:4-element IRS}(a) 4-element RIS with single connected
reconfigurable impedance network and (b) 4-element RIS with fully
connected reconfigurable impedance network.}
\end{figure}

\subsubsection{Single Connected Reconfigurable Impedance Network}

In this category, each port of the $N_{I}$ ports of the reconfigurable
impedance network are not connected to the other ports. An illustrative
example, for a 4-element RIS with a 4-port single connected reconfigurable
impedance network, is shown in Fig. \ref{fig:4-element IRS}(a). Generally,
for $n_{I}=1,\ldots,N_{I}$, the $n_{I}$th port is connected to ground
with a reconfigurable impedance $Z_{n_{I}}$, so that in total there
are $N_{I}$ reconfigurable impedance components in the network. Hence,
$\mathbf{Z}_{I}$ is a diagonal matrix given by $\mathbf{Z}_{I}=\mathrm{diag}\left(Z_{1},Z_{2},...,Z_{N_{I}}\right)$
and according to \eqref{eq:gammaI =00003D (ZI-Z0)(ZI+Z0)-1}, $\boldsymbol{\Theta}$
is also a diagonal matrix given by
\begin{equation}
\boldsymbol{\Theta}=\mathrm{diag}\left(\left[\boldsymbol{\Theta}\right]_{1,1},\left[\boldsymbol{\Theta}\right]_{2,2},...,\left[\boldsymbol{\Theta}\right]_{N_{I},N_{I}}\right),\label{eq:diag(gamma)}
\end{equation}
where the $\left(n_{I},n_{I}\right)$th entry of $\boldsymbol{\Theta}$,
denoted as $\left[\boldsymbol{\Theta}\right]_{n_{I},n_{I}}$, is the
reflection coefficient of the reconfigurable impedance $Z_{n_{I}}$,
i.e.
\begin{equation}
\left[\boldsymbol{\Theta}\right]_{n_{I},n_{I}}=\frac{Z_{n_{I}}-Z_{0}}{Z_{n_{I}}+Z_{0}},\label{eq:Fi Zi}
\end{equation}
so that we have an equivalent constraint that $\left|\left[\boldsymbol{\Theta}\right]_{n_{I},n_{I}}\right|\leq1$.
Furthermore, to increase the power scattered by RIS, $Z_{n_{I}}$
is purely reactive for $n_{I}=1,\ldots,N_{I}$, i.e. $Z_{n_{I}}=jX_{n_{I}}$
where $X_{n_{I}}$ denotes the reconfigurable reactance. Therefore
\begin{equation}
\left[\boldsymbol{\Theta}\right]_{n_{I},n_{I}}=\frac{jX_{n_{I}}-Z_{0}}{jX_{n_{I}}+Z_{0}}=e^{j\theta_{n_{I}}},\label{eq:ej_theta}
\end{equation}
where $0\leq\theta_{n_{I}}\leq2\pi$ denotes phase shift. Hence, we
have an equivalent unit modulus constraint of $\left|\left[\boldsymbol{\Theta}\right]_{n_{I},n_{I}}\right|=1$.
The single connected reconfigurable impedance network and the corresponding
constraints \eqref{eq:diag(gamma)}, \eqref{eq:ej_theta} have been
widely adopted in RIS aided wireless communication system designs
and optimizations \cite{2019_TWC_RIS_CHuang}-\cite{2020_WCL_IRS_DRL}.

\subsubsection{Fully Connected Reconfigurable Impedance Network}

In this paper, we propose a more general reconfigurable impedance
network, which is denoted as the fully connected reconfigurable impedance
network, to further improve the signal power received from RIS and
enhance the performance of RIS. In this category, each port of the
$N_{I}$ ports of the reconfigurable impedance network is connected
to other ports. An illustrative example, for a 4-element RIS with
a 4-port fully connected reconfigurable impedance network, is shown
in Fig. \ref{fig:4-element IRS}(b). Generally, for $n_{I}=1,\ldots,N_{I}$,
the $n_{I}$th port is connected to ground with a reconfigurable impedance
$Z_{n_{I}}$ and the $n_{I}$th port is connected to the $m_{I}$th
port, for $m_{I}=n_{I}+1,\ldots,N_{I}$, with a reconfigurable impedance
$Z_{n_{I},m_{I}}$, so that in total there are $N_{I}\left(N_{I}+1\right)/2$
reconfigurable impedance components in the network. Therefore, $\mathbf{Z}_{I}$
is a full matrix and, following \cite{ShanpuShen2016_TAP_Impedancematching},
$\mathbf{Z}_{I}$ can be obtained from the following relationship
\begin{equation}
\left[\mathbf{Z}_{I}^{-1}\right]_{n_{I},m_{I}}=\begin{cases}
-Z_{n_{I},m_{I}}^{-1} & ,n_{I}\neq m_{I}\\
Z_{n_{I}}^{-1}+\sum_{k\neq n_{I}}Z_{n_{I},k}^{-1} & ,n_{I}=m_{I}
\end{cases},\label{eq:ZI-1}
\end{equation}
where $Z_{n_{I},m_{I}}=Z_{m_{I},n_{I}}$ due to the symmetric $\mathbf{Z}_{I}$.
According to \eqref{eq:ZI-1}, we can implement an arbitrary impedance
matrix $\mathbf{Z}_{I}$ by selecting proper impedance $Z_{n_{I}}$
and $Z_{n_{I},m_{I}}$. Subsequently, $\boldsymbol{\Theta}$ can be
found by \eqref{eq:gammaI =00003D (ZI-Z0)(ZI+Z0)-1}. According to
network theory \cite{pozar2009microwave}, $\boldsymbol{\Theta}$
is a full matrix satisfying the constraints 
\begin{equation}
\boldsymbol{\Theta}=\boldsymbol{\Theta}^{T},\:\boldsymbol{\Theta}^{H}\boldsymbol{\Theta}\preceq\boldsymbol{\mathrm{I}}.
\end{equation}
Furthermore, to increase the power scattered by RIS, $Z_{n_{I}}$
and $Z_{n_{I},m_{I}}$ are purely reactive. That is $Z_{n_{I}}=jX_{n_{I}}$
and $Z_{n_{I},m_{I}}=jX_{n_{I},m_{I}}$ where $X_{n_{I}}$ and $X_{n_{I},m_{I}}$
denote the reconfigurable reactances, so that we have $\mathbf{Z}_{I}=j\mathbf{X}_{I}$
where $\mathbf{X}_{I}\in\mathbb{R}^{N_{I}\times N_{I}}$ denotes the
reactance matrix of the $N_{I}$-port reconfigurable impedance network
and $\mathbf{X}_{I}=\mathbf{X}_{I}^{T}$. Hence, $\boldsymbol{\Theta}$
is given by
\begin{equation}
\boldsymbol{\Theta}=\left(j\mathbf{X}_{I}+Z_{0}\mathbf{I}\right)^{-1}\left(j\mathbf{X}_{I}-Z_{0}\mathbf{I}\right),\label{eq:transform S to X}
\end{equation}
so that according to \cite{pozar2009microwave} we have equivalent
constraints 
\begin{equation}
\boldsymbol{\Theta}=\boldsymbol{\Theta}^{T},\:\boldsymbol{\Theta}^{H}\boldsymbol{\Theta}=\boldsymbol{\mathrm{I}},\label{eq:gamma unitary constraint}
\end{equation}
which shows that $\boldsymbol{\Theta}$ is a complex symmetric unitary
matrix. The single connected reconfigurable impedance network \eqref{eq:diag(gamma)},
\eqref{eq:ej_theta} is a special case of the fully connected reconfigurable
impedance network \eqref{eq:gamma unitary constraint}, so that the
fully connected reconfigurable impedance network is more general and
is expected to provide better RIS performance.

\begin{figure}[t]
\begin{centering}
\includegraphics[width=8.5cm]{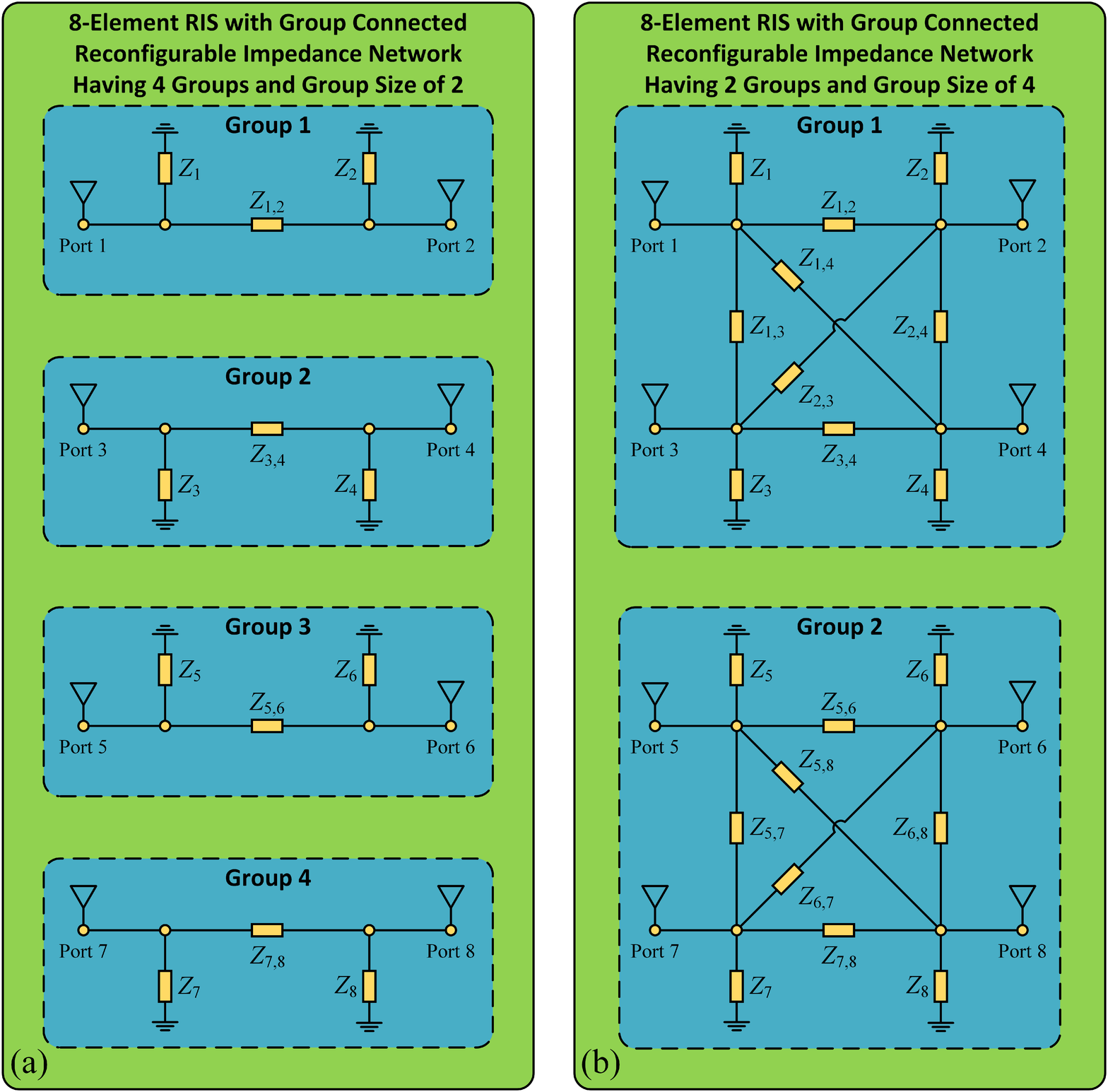}
\par\end{centering}
\caption{\label{fig:8-element }8-element RIS with group connected reconfigurable
impedance network (a) having 4 groups with group size of 2 and (b)
having 2 groups with group size of 4.}
\end{figure}

\subsubsection{Group Connected Reconfigurable Impedance Network}

For the fully connected reconfigurable impedance network, when $N_{I}$
becomes large, the number of reconfigurable impedance components becomes
huge (increasing quadratically with $N_{I}$) and the circuit topology
will become intricate. This limits its practical use and therefore
we also propose a group connected reconfigurable impedance network
to achieve a good tradeoff between performance enhancement and complexity.
Two illustrative examples, for an 8-element RIS with a group connected
reconfigurable impedance network having 4 groups and 2 groups, are
shown in Fig. \ref{fig:8-element }(a) and (b), respectively. In Fig.
\ref{fig:8-element }(a), the 8 elements in RIS are divided into 4
groups and each group has 2 elements and uses a 2-port fully connected
reconfigurable impedance network. In Fig. \ref{fig:8-element }(b),
the 8 elements in RIS are divided into 2 groups and each group has
4 elements and uses a 4-port fully connect reconfigurable impedance
network. Generally, for $N_{I}$-element RIS, we can divide it into
$G$ groups with each group having $N_{G}=\frac{N_{I}}{G}$ elements.
We refer to $N_{G}$ as the group size. For the $g$th group, a $N_{G}$-port
fully connected reconfigurable impedance network with impedance matrix
of $\mathbf{Z}_{I,g}\in\mathbb{C}^{N_{G}\times N_{G}}$ is used. Therefore,
$\mathbf{Z}_{I}$ is a block diagonal matrix given by
\begin{equation}
\mathbf{Z}_{I}=\mathrm{diag}\left(\mathbf{Z}_{I,1},\mathbf{Z}_{I,2},...,\mathbf{Z}_{I,G}\right).
\end{equation}
Subsequently $\boldsymbol{\Theta}$ can be found by \eqref{eq:gammaI =00003D (ZI-Z0)(ZI+Z0)-1}.
According to \eqref{eq:gammaI =00003D (ZI-Z0)(ZI+Z0)-1} and \cite{pozar2009microwave},
$\boldsymbol{\Theta}$ is a block diagonal matrix satisfying the constraints
\begin{equation}
\boldsymbol{\Theta}=\mathrm{diag}\left(\boldsymbol{\Theta}_{1},\boldsymbol{\Theta}_{2},...,\boldsymbol{\Theta}_{G}\right),\label{eq:diag(group)}
\end{equation}
\begin{equation}
\boldsymbol{\Theta}_{g}=\boldsymbol{\Theta}_{g}^{T},\:\boldsymbol{\Theta}_{g}^{H}\boldsymbol{\Theta}_{g}\preceq\boldsymbol{\mathrm{I}},\:\forall g.
\end{equation}
Furthermore, to increase the power scattered by RIS, $\mathbf{Z}_{I,g}$
are purely reactive, i.e. $\mathbf{Z}_{I,g}=j\mathbf{X}_{I,g}$ where
$\mathbf{X}_{I,g}\in\mathbb{R}^{N_{G}\times N_{G}}$ denotes the reactance
matrix of the $N_{G}$-port reconfigurable impedance network and $\mathbf{X}_{I,g}=\mathbf{X}_{I,g}^{T}$.
Accordingly, $\boldsymbol{\Theta}_{g}$ can be found as
\begin{equation}
\boldsymbol{\Theta}_{g}=\left(j\mathbf{X}_{I,g}+Z_{0}\mathbf{I}\right)^{-1}\left(j\mathbf{X}_{I,g}-Z_{0}\mathbf{I}\right),\label{eq:transform S to X BLOCK}
\end{equation}
so that according to \cite{pozar2009microwave} we have equivalent
constraints 
\begin{equation}
\boldsymbol{\Theta}_{g}=\boldsymbol{\Theta}_{g}^{T},\:\boldsymbol{\Theta}_{g}^{H}\boldsymbol{\Theta}_{g}=\boldsymbol{\mathrm{I}},\:\forall g.\label{eq:group gamma unitary constraint}
\end{equation}
Therefore, $\boldsymbol{\Theta}$ is a block diagonal matrix with
each block being a complex symmetric unitary matrix. For an $N_{I}$-port
group connected reconfigurable impedance network, there are in total
$N_{I}\left(N_{G}+1\right)/2$ reconfigurable impedance components.
When the group size $N_{G}=1$, it becomes the single connected reconfigurable
impedance network. When the group size $N_{G}=N_{I}$, it becomes
the fully connected reconfigurable impedance network.

Comparisons between the single connected, fully connected, and group
connected reconfigurable impedance networks will be shown in the following
sections.

\section{RIS Aided Communication Model}

We have analyzed the fundamental relationships between the incident
and reflected waves of the antennas at the transmitter, receiver,
and RIS. In this section, we establish the RIS aided communication
model based on these relationships.

\subsection{General RIS Aided Communication Model}

We first consider a general RIS aided communication model. Combining
\eqref{eq:atbt}, \eqref{eq:arbr}, and \eqref{eq:aibi}, we can relate
$\mathbf{a}$ and $\mathbf{b}$ in a compact form as
\begin{equation}
\mathbf{a}=\mathbf{b}_{s}+\mathbf{\Gamma}\mathbf{b},\label{eq:a =00003D bs+Fb}
\end{equation}
where $\mathbf{b}_{s}$ and $\mathbf{\Gamma}$ are respectively given
by
\begin{equation}
\mathbf{b}_{s}=\left[\begin{array}{c}
\mathbf{b}_{s,T}\\
\mathbf{0}\\
\mathbf{0}
\end{array}\right],\:\mathbf{\Gamma}=\left[\begin{array}{ccc}
\mathbf{\Gamma}_{T} & \mathbf{0} & \mathbf{0}\\
\mathbf{0} & \boldsymbol{\Theta} & \mathbf{0}\\
\mathbf{0} & \mathbf{0} & \mathbf{\Gamma}_{R}
\end{array}\right].\label{eq:gamma_and_theta}
\end{equation}
Substituting \eqref{eq:b=00003DSa} into \eqref{eq:a =00003D bs+Fb},
we have that
\begin{equation}
\mathbf{\mathbf{b}}=\mathbf{S}\left(\mathbf{I}-\mathbf{\Gamma}\mathbf{S}\right)^{-1}\mathbf{b}_{s}.
\end{equation}
We define $\mathbf{T}\triangleq\mathbf{S}\left(\mathbf{I}-\mathbf{\Gamma}\mathbf{S}\right)^{-1}\in\mathbb{C}^{N\times N}$
and partition $\mathbf{T}$ as
\begin{equation}
\mathbf{T}=\left[\begin{array}{ccc}
\mathbf{T}_{TT} & \mathbf{T}_{TI} & \mathbf{T}_{TR}\\
\mathbf{T}_{IT} & \mathbf{T}_{II} & \mathbf{T}_{IR}\\
\mathbf{T}_{RT} & \mathbf{T}_{RI} & \mathbf{T}_{RR}
\end{array}\right],
\end{equation}
so that we can find $\mathbf{\mathbf{b}}_{T}$ and $\mathbf{\mathbf{b}}_{R}$
as
\begin{equation}
\mathbf{\mathbf{b}}_{T}=\mathbf{T}_{TT}\mathbf{b}_{s,T},\:\mathbf{\mathbf{b}}_{R}=\mathbf{T}_{RT}\mathbf{b}_{s,T}.\label{eq:bt}
\end{equation}

We define the voltage vector at the transmitter as $\mathbf{v}_{T}\triangleq\left[v_{T,1},v_{T,2},\ldots,v_{T,N_{T}}\right]^{T}\in\mathbb{C}^{N_{T}\times1}$
where $v_{T,n_{T}}$ refers to the voltage across the $n_{T}$th transmit
antenna. We also define the voltage vector at the receiver as $\mathbf{v}_{R}\triangleq\left[v_{R,1},v_{R,2},\ldots,v_{R,N_{R}}\right]^{T}\in\mathbb{C}^{N_{R}\times1}$
where $v_{R,n_{R}}$ refers to the voltage across the $n_{R}$th receive
antenna. With the incident and reflected waves, we can find $\mathbf{v}_{T}$
and $\mathbf{v}_{R}$ as
\begin{equation}
\mathbf{v}_{T}=\mathbf{a}_{T}+\mathbf{b}_{T},\:\mathbf{v}_{R}=\mathbf{a}_{R}+\mathbf{b}_{R}.\label{eq:vr}
\end{equation}
More details about the relationship between the voltages and the incident
and reflected waves of an $N$-port network can be found in the Appendix.
Utilizing \eqref{eq:atbt}, \eqref{eq:arbr}, and \eqref{eq:bt}-\eqref{eq:vr},
we can relate $\mathbf{v}_{T}$ and $\mathbf{v}_{R}$ by
\begin{equation}
\mathbf{v}_{R}=\left(\mathbf{\Gamma}_{R}+\mathbf{I}\right)\mathbf{T}_{RT}\left(\mathbf{I}+\mathbf{\Gamma}_{T}\mathbf{T}_{TT}+\mathbf{T}_{TT}\right)^{-1}\mathbf{v}_{T}.
\end{equation}
Defining $\mathbf{v}_{T}$ as the transmit signal $\mathbf{x}$ and
$\mathbf{v}_{R}$ as the receive signal $\mathbf{y}$, we can find
the channel matrix of the RIS aided wireless communication system
as
\begin{equation}
\mathbf{H}=\left(\mathbf{\Gamma}_{R}+\mathbf{I}\right)\mathbf{T}_{RT}\left(\mathbf{I}+\mathbf{\Gamma}_{T}\mathbf{T}_{TT}+\mathbf{T}_{TT}\right)^{-1},\label{eq:H general}
\end{equation}
so that we have $\mathbf{y}=\mathbf{H}\mathbf{x}$ (ignoring the additive
white Gaussian noise (AWGN) at the receiver). According to \eqref{eq:gamma_and_theta}
and $\mathbf{T}\triangleq\mathbf{S}\left(\mathbf{I}-\mathbf{\Gamma}\mathbf{S}\right)^{-1}$,
the submatrice $\mathbf{T}_{TT}$ and $\mathbf{T}_{RT}$ are functions
of $\boldsymbol{\Theta}$, denoted as $\mathbf{T}_{TT}\left(\boldsymbol{\Theta}\right)$
and $\mathbf{T}_{RT}\left(\boldsymbol{\Theta}\right)$, so that from
\eqref{eq:H general} the channel matrix $\mathbf{H}$ is also a function
of $\boldsymbol{\Theta}$, denoted as $\mathbf{H}\left(\boldsymbol{\Theta}\right)$.
Hence, we can optimize $\boldsymbol{\Theta}$ to intelligently control
the channel $\mathbf{H}\left(\boldsymbol{\Theta}\right)$ and enhance
the wireless system performance.

The general communication model \eqref{eq:H general} includes the
effects of impedance mismatching and mutual coupling at the transmitter,
RIS, and receiver. However, generally it is difficult to find expressions
for $\mathbf{T}_{TT}\left(\boldsymbol{\Theta}\right)$ and $\mathbf{T}_{RT}\left(\boldsymbol{\Theta}\right)$
due to the matrix inversion operation. Subsequently, it is difficult
to find the expressions of $\mathbf{H}\left(\boldsymbol{\Theta}\right)$,
which makes it difficult to obtain insight into the role of RIS in
the communication model and to optimize $\boldsymbol{\Theta}$ of
RIS. Considering this issue, in the following, we consider a special
case to simplify the expression of $\mathbf{\mathbf{H}\left(\boldsymbol{\Theta}\right)}$.

\subsection{RIS Aided Communication Model with Perfect Matching and No Mutual
Coupling}

We consider a special case that assumes the antenna arrays at the
transmitter, RIS, and receiver are perfectly matched and have no mutual
coupling, i.e. $\mathbf{S}_{TT}=\mathbf{0}$, $\mathbf{S}_{II}=\mathbf{0}$,
and $\mathbf{S}_{RR}=\mathbf{0}$. In practice, this assumption can
be approximately achieved by individually matching each antenna to
the reference impedance $Z_{0}$ and keeping the antenna spacing larger
than half-wavelength. It is also assumed that the source impedance
at the transmitter $Z_{T,n_{T}}$ and the load impedance at the receiver
$Z_{R,n_{R}}$ are all reference impedances $Z_{0}$ so that $\mathbf{\Gamma}_{T}=\mathbf{0}$
and $\mathbf{\Gamma}_{R}=\mathbf{0}$. With these two assumptions,
we can simplify
\begin{equation}
\mathbf{S}=\left[\begin{array}{ccc}
\mathbf{0} & \mathbf{S}_{TI} & \mathbf{S}_{TR}\\
\mathbf{S}_{IT} & \mathbf{0} & \mathbf{S}_{IR}\\
\mathbf{S}_{RT} & \mathbf{S}_{RI} & \mathbf{0}
\end{array}\right],\:\mathbf{\Gamma}=\left[\begin{array}{ccc}
\mathbf{0} & \mathbf{0} & \mathbf{0}\\
\mathbf{0} & \boldsymbol{\Theta} & \mathbf{0}\\
\mathbf{0} & \mathbf{0} & \mathbf{0}
\end{array}\right],\label{eq:simplified S and Gamma}
\end{equation}
so that accordingly we can simplify $\left(\mathbf{I}-\mathbf{\Gamma}\mathbf{S}\right)^{-1}$
as
\begin{equation}
\left(\mathbf{I}-\mathbf{\Gamma}\mathbf{S}\right)^{-1}=\left[\begin{array}{ccc}
\mathbf{I} & \mathbf{0} & \mathbf{0}\\
\boldsymbol{\Theta}\mathbf{S}_{IT} & \mathbf{I} & \boldsymbol{\Theta}\mathbf{S}_{IR}\\
\mathbf{0} & \mathbf{0} & \mathbf{I}
\end{array}\right].\label{eq:(I-GammaS)-1}
\end{equation}
Making use of \eqref{eq:simplified S and Gamma} and \eqref{eq:(I-GammaS)-1},
we can simplify the expression of $\mathbf{T}=\mathbf{S}\left(\mathbf{I}-\mathbf{\Gamma}\mathbf{S}\right)^{-1}$
and then write that
\begin{align}
\mathbf{T}_{TT} & =\mathbf{S}_{TI}\boldsymbol{\Theta}\mathbf{S}_{IT},\label{eq:Ttt}\\
\mathbf{T}_{RT} & =\mathbf{S}_{RT}+\mathbf{S}_{RI}\boldsymbol{\Theta}\mathbf{S}_{IT}.\label{eq:Trt}
\end{align}
Substituting \eqref{eq:Ttt} and \eqref{eq:Trt} into \eqref{eq:H general}
and making use of $\mathbf{\Gamma}_{T}=\mathbf{0}$ and $\mathbf{\Gamma}_{R}=\mathbf{0}$,
we can simplify the channel matrix $\mathbf{H}$ as
\begin{equation}
\mathbf{H}=\left(\mathbf{S}_{RT}+\mathbf{S}_{RI}\boldsymbol{\Theta}\mathbf{S}_{IT}\right)\left(\mathbf{I}+\mathbf{S}_{TI}\boldsymbol{\Theta}\mathbf{S}_{IT}\right)^{-1}.
\end{equation}
The term $\mathbf{S}_{TI}\boldsymbol{\Theta}\mathbf{S}_{IT}$ refers
to the second order reflections between the transmitter and RIS and
back to the transmitter. However, in most applications, there is no
need to consider these second order reflections because the power
of the second reflection $\mathbf{S}_{TI}\boldsymbol{\Theta}\mathbf{S}_{IT}$
is extremely small, and is proportional to the square of the pathloss
between the transmitter to RIS. Hence, in most applications, we can
approximate $\left(\mathbf{I}+\mathbf{S}_{TI}\boldsymbol{\Theta}\mathbf{S}_{IT}\right)^{-1}$
as $\mathbf{I}$ without affecting the accuracy, and then simplify
$\mathbf{H}$ as
\begin{equation}
\mathbf{H}=\mathbf{S}_{RT}+\mathbf{S}_{RI}\boldsymbol{\Theta}\mathbf{S}_{IT}.\label{eq: Channel in S}
\end{equation}
The transmission scattering matrices $\mathbf{S}_{RT}$, $\mathbf{S}_{IT}$,
and $\mathbf{S}_{RI}$ are equivalently the channel matrices from
the transmitter to receiver, from the transmitter to RIS, and from
the RIS to receiver, respectively. To show the equivalence, we consider
the impedance matrix of the $N$-port network
\begin{equation}
\mathbf{Z}=\left[\begin{array}{ccc}
\mathbf{Z}_{TT} & \mathbf{Z}_{TI} & \mathbf{Z}_{TR}\\
\mathbf{Z}_{IT} & \mathbf{Z}_{II} & \mathbf{Z}_{IR}\\
\mathbf{Z}_{RT} & \mathbf{Z}_{RI} & \mathbf{Z}_{RR}
\end{array}\right].
\end{equation}
Since we assume perfect matching and no mutual coupling, we have $\mathbf{Z}_{TT}=\mathbf{Z}_{II}=\mathbf{Z}_{RR}=Z_{0}\mathbf{I}$
and following \cite{pozar2009microwave} we can derive
\begin{equation}
\mathbf{S}_{ij}=\frac{\mathbf{Z}_{ij}}{2Z_{0}},\label{eq:sij=00003Dzij/2Z0}
\end{equation}
for $ij\in\left\{ RT,RI,IT\right\} $. We take $\mathbf{Z}_{RT}$
as an example to see the details. The $\left(n_{R},n_{T}\right)$th
entry of $\mathbf{Z}_{RT}$, denoted as $\left[\mathbf{Z}_{RT}\right]_{n_{R},n_{T}}$,
refers to the trans-impedance between the $n_{T}$th transmit antenna
and $n_{R}$th receive antenna. To find $\left[\mathbf{Z}_{RT}\right]_{n_{R},n_{T}}$,
we excite the $n_{T}$th transmit antenna with current $i_{T,n_{T}}$
and keep all the other antennas open circuited, and then measure the
open-circuit voltage $v_{R,n_{R}}^{\mathrm{open}}$ at the $n_{R}$th
receive antenna. Using the multipath propagation based model, we have
that
\begin{equation}
v_{R,n_{R}}^{\mathrm{open}}=\underset{\left[\mathbf{Z}_{RT}\right]_{n_{R},n_{T}}}{\underbrace{\left(\sum_{l=1}^{\mathscr{L}_{RT}}e_{R,n_{R}}\left(\Omega_{R,l}\right)\beta_{RT,l}e_{T,n_{T}}\left(\Omega_{T,l}\right)\right)}}i_{T,n_{T}},\label{eq:multiple path prog}
\end{equation}
where $e_{T,n_{T}}\left(\cdot\right)$ and $e_{R,n_{R}}\left(\cdot\right)$
denote the open-circuit radiation pattern of the $n_{T}$th transmit
antenna and the $n_{R}$th receive antenna, respectively. We assume
a channel with $\mathcal{\mathscr{L}}_{RT}$ paths for propagation
from transmitter to receiver with the $l$th path characterized by
departure and arrival angles $\Omega_{T,l}$ and $\Omega_{R,l}$,
respectively, and a complex channel gain $\beta_{RT,l}$. Therefore,
from \eqref{eq:sij=00003Dzij/2Z0} and \eqref{eq:multiple path prog},
we show that $\mathbf{S}_{RT}$ is equivalently the channel matrix
from the transmitter to receiver. Similarly, we can show that $\mathbf{S}_{IT}$
and $\mathbf{S}_{RI}$ are the channel matrices from the transmitter
to RIS and from the RIS to receiver, respectively. We use auxiliary
notations $\mathbf{H}_{RT}=\mathbf{S}_{RT}$, $\mathbf{H}_{IT}=\mathbf{S}_{IT}$,
and $\mathbf{H}_{RI}=\mathbf{S}_{RI}$ to facilitate understanding,
so that we can rewrite \eqref{eq: Channel in S} as 
\begin{equation}
\mathbf{H}=\mathbf{H}_{RT}+\mathbf{H}_{RI}\boldsymbol{\Theta}\mathbf{H}_{IT}.\label{eq:perfectmatchign H}
\end{equation}
Furthermore, assuming there are LoS and non-LoS (NLoS) paths in \eqref{eq:multiple path prog},
we can model $\mathbf{H}_{RT}$, $\mathbf{H}_{IT}$, and $\mathbf{H}_{RI}$
as Rician fading, i.e.
\begin{equation}
\mathbf{H}_{ij}=\sqrt{L_{ij}}\left(\sqrt{\frac{K_{ij}}{1+K_{ij}}}\mathbf{H}_{ij}^{\mathrm{LoS}}+\sqrt{\frac{1}{1+K_{ij}}}\mathbf{H}_{ij}^{\mathrm{NLoS}}\right),\label{eq:Rician Channel Model}
\end{equation}
for $ij\in\left\{ RT,RI,IT\right\} $ where $L_{ij}$ refers to the
pathloss, $K_{ij}$ refers to the Rician factor, $\mathbf{H}_{ij}^{\mathrm{LoS}}$
and $\mathbf{H}_{ij}^{\mathrm{NLoS}}$ represent the small-scale LoS
and NLoS (Rayleigh fading) components, respectively.

The simplified $\mathbf{H}$ \eqref{eq:perfectmatchign H} is a linear
function of $\boldsymbol{\Theta}$ and together with
\begin{enumerate}
\item the single connected reconfigurable impedance network which satisfies
the constraint that $\boldsymbol{\Theta}=\mathrm{diag}\left(e^{j\theta_{1}},e^{j\theta_{2}},...,e^{j\theta_{N_{I}}}\right)$,
\item the fully connected reconfigurable impedance network which satisfies
the constraints that $\boldsymbol{\Theta}=\boldsymbol{\Theta}^{T}$,
$\boldsymbol{\Theta}^{H}\boldsymbol{\Theta}=\boldsymbol{\mathrm{I}}$,
\item the group connected reconfigurable impedance network which satisfies
the constraints that $\boldsymbol{\Theta}=\mathrm{diag}\left(\boldsymbol{\Theta}_{1},\boldsymbol{\Theta}_{2},...,\boldsymbol{\Theta}_{G}\right)$,
$\boldsymbol{\Theta}_{g}=\boldsymbol{\Theta}_{g}^{T}$, $\boldsymbol{\Theta}_{g}^{H}\boldsymbol{\Theta}_{g}=\boldsymbol{\mathrm{I}}$,
$\forall g$,
\end{enumerate}
make up our proposed RIS aided communication model. Based on this
model, we can optimize $\boldsymbol{\Theta}$ to intelligently control
the channel $\mathbf{H}\left(\boldsymbol{\Theta}\right)$ and enhance
the wireless system performance. Note that, the conventional RIS aided
communication model used in \cite{2019_TWC_RIS_CHuang}-\cite{2020_WCL_IRS_DRL}
is a special case of our proposed model that corresponds to the single
connected reconfigurable impedance network. Importantly, in sharp
contrast with the conventional single connected architecture \cite{2019_TWC_RIS_CHuang}-\cite{2020_WCL_IRS_DRL}
that only adjusts the phases of the impinging waves using a diagonal
scattering matrix, our proposed group and fully connected architectures
enable scattering matrices to be block diagonal or full and can consequently
adjust not only the phases but also the magnitudes of the impinging
waves. This leads to significant performance gains in fading channels
as it will appear in Sections IV and V.
\begin{rem}
It is worthwhile to clarify the differences between our proposed model
with the RIS aided communication model proposed in recent work \cite{2020_ArXiv_RIS_MutualImpedance}.
There are four differences. \textit{First}, compared with using impedance
parameter in \cite{2020_ArXiv_RIS_MutualImpedance}, we have found
it is more natural to use the reflection coefficient and scattering
parameter to account for the scattering mechanism of RIS and derive
the RIS aided communication model. \textit{Second}, our general RIS
aided communication model \eqref{eq:H general} includes the effects
of impedance mismatching and mutual coupling at the transmitter, RIS,
and receiver, which is more general than \cite{2020_ArXiv_RIS_MutualImpedance}.
\textit{Third}, we clearly explain the physical significance of the
phase shifts and the unit modulus constraint. \textit{Fourth}, we
go beyond the single connected reconfigurable impedance network (the
main focus in \cite{2020_ArXiv_RIS_MutualImpedance}) and propose
more general fully connected and group connected reconfigurable impedance
networks.
\end{rem}

\section{Scaling Law}

In order to obtain insights into the fundamental limits of single,
fully, and group connected reconfigurable impedance networks in RIS,
we quantify how the received signal power scales as a function of
the number of RIS elements $N_{I}$. For simplicity, we consider a
SISO RIS aided system ($N_{T}=1$, $N_{R}=1$) with perfect matching
and no mutual coupling in the following. The transmit signal is $x\in\mathbb{C}$
with $\mathrm{E}\left[\left|x\right|^{2}\right]=P_{T}$. According
to \eqref{eq:perfectmatchign H}, the received signal $y$ can be
expressed as
\begin{equation}
y=\left(h_{RT}+\mathbf{h}_{RI}\boldsymbol{\Theta}\mathbf{h}_{IT}\right)\mathbf{x}+n,
\end{equation}
where $h_{RT}\in\mathbb{C}$, $\mathbf{h}_{IT}\in\mathbb{C}^{N_{I}\times1}$,
and $\mathbf{h}_{RI}\in\mathbb{C}^{1\times N_{I}}$ denote the channel
from the transmitter to the receiver, from the transmitter to the
RIS, and from the RIS to the receiver, respectively, and $n$ is the
AWGN. For simplicity, we assume the transmit power $P_{T}=1$ and
omit the direct channel $h_{RT}$, so that we can express the received
signal power as $P_{R}=\left|\mathbf{h}_{RI}\boldsymbol{\Theta}\mathbf{h}_{IT}\right|^{2}$.

\subsection{Single Connected Reconfigurable Impedance Network}

For the single connected reconfigurable impedance network \eqref{eq:diag(gamma)},
\eqref{eq:ej_theta}, it is obvious that the optimal $\boldsymbol{\Theta}^{\star}$
is
\begin{align}
\boldsymbol{\Theta}^{\star} & =\mathrm{diag}\left(e^{j\theta_{1}^{\star}},e^{j\theta_{2}^{\star}},...,e^{j\theta_{N_{I}}^{\star}}\right),\label{eq:optimal single connected-1}\\
\theta_{n_{I}}^{\star} & =-\arg\left(\left[\mathbf{h}_{RI}\right]_{n_{I}}\left[\mathbf{h}_{IT}\right]_{n_{I}}\right),\forall n_{I},\label{eq:optimal single connected-2}
\end{align}
which achieves the maximum received signal power
\begin{equation}
P_{R}^{\mathrm{Single}}=\left(\sum_{n_{I}=1}^{N_{I}}\left|\left[\mathbf{h}_{RI}\right]_{n_{I}}\left[\mathbf{h}_{IT}\right]_{n_{I}}\right|\right)^{2}.\label{eq:maximum Power single}
\end{equation}

\subsection{Fully Connected Reconfigurable Impedance Network}

For the fully connected reconfigurable impedance network \eqref{eq:gamma unitary constraint},
using the Cauchy-Schwarz inequality and that $\boldsymbol{\Theta}^{H}\boldsymbol{\Theta}=\boldsymbol{\mathrm{I}}$,
we can find an upper bound for the maximum received signal power $P_{R}^{\mathrm{Fully}}$
as
\begin{equation}
P_{R}^{\mathrm{Fully}}\leq\bar{P}_{R}^{\mathrm{Fully}}=\left\Vert \mathbf{h}_{RI}\right\Vert ^{2}\left\Vert \mathbf{h}_{IT}\right\Vert ^{2}.\label{eq:upper bound of fully}
\end{equation}
The key to achieve the upper bound $\bar{P}_{R}^{\mathrm{Fully}}$
is that we need to find a complex symmetric unitary matrix $\boldsymbol{\Theta}$
satisfying 
\begin{equation}
\frac{\mathbf{h}_{RI}^{H}}{\left\Vert \mathbf{h}_{RI}\right\Vert }=\boldsymbol{\Theta}\frac{\mathbf{h}_{IT}}{\left\Vert \mathbf{h}_{IT}\right\Vert }.\label{eq:optimial fully}
\end{equation}
However, it is difficult to derive a closed-form solution for the
optimal $\boldsymbol{\Theta}^{\star}$ which satisfies the equation
\eqref{eq:optimial fully} and achieves the upper bound. Hence, we
directly optimize $\boldsymbol{\Theta}$ to approach the upper bound
using the quasi-Newton method as detailed in the next section. Numerical
results using the Monte Carlo method confirms that the upper bound
\eqref{eq:upper bound of fully} is tight.

It is also worthwhile to compare the maximum received signal power
of the single connected and fully connected reconfigurable impedance
networks.  From \eqref{eq:maximum Power single} and \eqref{eq:upper bound of fully},
using Cauchy-Schwarz inequality we can deduce that 
\begin{equation}
P_{R}^{\mathrm{Single}}\leq\bar{P}_{R}^{\mathrm{Fully}},\label{eq:single<fully}
\end{equation}
and the equality is achieved if and only if 
\begin{equation}
\left|\left[\mathbf{h}_{RI}\right]_{n_{I}}\right|=\alpha\left|\left[\mathbf{h}_{IT}\right]_{n_{I}}\right|,\:\forall n_{I},\label{eq:equal conditions}
\end{equation}
where $\alpha$ can be any positive scalar. In other words, when the
channel gains (the modulus) of $\mathbf{h}_{RI}$ and $\mathbf{h}_{IT}$
are linearly independent, the fully connected reconfigurable impedance
network can achieve a higher received signal power than the single
connected case.

We also provide physical explanations to account for the better performance
of the fully connected case. For the single connected case, each port
of the reconfigurable impedance network is not connected to other
ports. As a result, only the phase of the elements of the vector $\boldsymbol{\Theta}\mathbf{h}_{IT}$
can be adjusted. Therefore the best that RIS can achieve is to make
the two channel vectors $\mathbf{h}_{RI}$ and $\mathbf{h}_{IT}$
element-wise in phase. However, for the fully connected case, each
port of the reconfigurable impedance network is connected to each
other. As a result, the phase and magnitude of the elements of the
vector $\boldsymbol{\Theta}\mathbf{h}_{IT}$ can be jointly adjusted
so that the RIS can align the two channel vectors $\mathbf{h}_{RI}$
and $\mathbf{h}_{IT}$ in the same direction to achieve a better performance.
Intuitively speaking, the single connected case is analogous to the
equal-gain combining while the fully connected case is analogous to
the maximum ratio combining. Namely, instead of adjusting only the
phase of the impinging wave as in the single connected architecture,
the fully connected architecture can adjust not only the phases but
also the magnitudes of the impinging waves.

\subsection{Group Connected Reconfigurable Impedance Network}

For the group connected reconfigurable impedance network \eqref{eq:diag(group)},
\eqref{eq:group gamma unitary constraint}, we can rewrite the received
signal power as
\begin{equation}
P_{R}=\left|\sum_{g=1}^{G}\mathbf{h}_{RI,g}\boldsymbol{\Theta}_{g}\mathbf{h}_{IT,g}\right|^{2},
\end{equation}
where $\mathbf{h}_{RI}=\left[\mathbf{h}_{RI,1},\mathbf{h}_{RI,2},\ldots,\mathbf{h}_{RI,G}\right]$
with $\mathbf{h}_{RI,g}\in\mathbb{C}^{1\times N_{G}}$ and $\mathbf{h}_{IT}=\left[\mathbf{h}_{IT,1},\mathbf{h}_{IT,2},\ldots,\mathbf{h}_{IT,G}\right]^{T}$
with $\mathbf{h}_{IT,g}\in\mathbb{C}^{N_{G}\times1}$. Using the triangle
inequality, Cauchy-Schwarz inequality, and that $\boldsymbol{\Theta}_{g}^{H}\boldsymbol{\Theta}_{g}=\boldsymbol{\mathrm{I}}$
$\forall g$, we can find an upper bound for the maximum received
signal power $P_{R}^{\mathrm{Group}}$ as
\begin{equation}
P_{R}^{\mathrm{Group}}\leq\bar{P}_{R}^{\mathrm{Group}}=\left(\sum_{g=1}^{G}\left\Vert \mathbf{h}_{RI,g}\right\Vert \left\Vert \mathbf{h}_{IT,g}\right\Vert \right)^{2}.\label{eq:upper bound in PR}
\end{equation}
The key to achieve the upper bound $\bar{P}_{R}^{\mathrm{Group}}$
is that we need to find a complex symmetric unitary matrix $\boldsymbol{\Theta}_{g}$
satisfying 
\begin{equation}
\frac{\mathbf{h}_{RI,g}^{H}}{\left\Vert \mathbf{h}_{RI,g}\right\Vert }=\boldsymbol{\Theta}_{g}\frac{\mathbf{h}_{IT,g}}{\left\Vert \mathbf{h}_{IT,g}\right\Vert },\forall g.
\end{equation}
However, it is difficult to derive a closed-form solution for the
optimal $\boldsymbol{\Theta}_{g}^{\star}$ achieving the upper bound.
Similar to the fully connected case, we directly optimize $\boldsymbol{\Theta}_{g}$
$\forall g$ to approach the upper bound using the quasi-Newton method
and numerical results using the Monte Carlo method confirm that the
upper bound \eqref{eq:upper bound in PR} is tight.

The group connected case can be viewed as a tradeoff between the single
connected and fully connected cases so that it is straightforward
to show that $P_{R}^{\mathrm{Single}}\leq\bar{P}_{R}^{\mathrm{Group}}\leq\bar{P}_{R}^{\mathrm{Fully}}$.
In the next subsections, we investigate $P_{R}^{\mathrm{Single}}$,
$P_{R}^{\mathrm{Fully}}$, and $P_{R}^{\mathrm{Group}}$ with LoS
and Rayleigh fading channels.

\subsection{Line-of-Sight Channel}

Assuming $\mathbf{h}_{RI}$ and $\mathbf{h}_{IT}$ are both LoS channels,
we have that $\mathbf{h}_{RI}=\left[e^{j\phi_{1}},\ldots,e^{j\phi_{N_{I}}}\right]$
and $\mathbf{h}_{IT}=\left[e^{j\psi_{1}},\ldots,e^{j\psi_{N_{I}}}\right]^{T}$.
In this case, it is obvious that the optimal $\boldsymbol{\Theta}^{\star}$
for the single, fully, and group connected reconfigurable impedance
networks are the same and given by 
\begin{equation}
\boldsymbol{\Theta}^{\star}=\mathrm{diag}\left(e^{-j\left(\phi_{1}+\psi_{1}\right)},...,e^{-j\left(\phi_{N_{I}}+\psi_{N_{I}}\right)}\right).
\end{equation}
Therefore, the single, fully, and group connected reconfigurable impedance
networks have the same performance with LoS channel such that 
\begin{equation}
P_{R}^{\mathrm{Single}}=P_{R}^{\mathrm{Fully}}=P_{R}^{\mathrm{Group}}=N_{I}^{2}.\label{eq:scaling law-LOS}
\end{equation}
This is consistent with that the equality in \eqref{eq:single<fully}
that can be achieved when \eqref{eq:equal conditions} is satisfied.

\subsection{Rayleigh Fading Channel}

Assuming $\mathbf{h}_{RI}$ and $\mathbf{h}_{IT}$ are independent
and identically distributed (i.i.d.) Rayleigh fading channels, we
have $\mathbf{h}_{RI}\sim\mathcal{CN}\left(\boldsymbol{0},\mathbf{I}\right)$
and $\mathbf{h}_{IT}\sim\mathcal{CN}\left(\boldsymbol{0},\mathbf{I}\right)$.
We first consider the group connected case. We rewrite $\bar{P}_{R}^{\mathrm{Group}}$
as
\begin{align}
\bar{P}_{R}^{\mathrm{Group}} & =\sum_{g=1}^{G}\left\Vert \mathbf{h}_{RI,g}\right\Vert ^{2}\left\Vert \mathbf{h}_{IT,g}\right\Vert ^{2}\nonumber \\
 & +\sum_{g_{1}\neq g_{2}}\left\Vert \mathbf{h}_{RI,g_{1}}\right\Vert \left\Vert \mathbf{h}_{IT,g_{1}}\right\Vert \left\Vert \mathbf{h}_{RI,g_{2}}\right\Vert \left\Vert \mathbf{h}_{IT,g_{2}}\right\Vert ,
\end{align}
Taking the expectation and making use of the i.i.d. Rayleigh fading
assumption of $\mathbf{h}_{RI}$ and $\mathbf{h}_{IT}$, we can find
the average $\bar{P}_{R}^{\mathrm{Group}}$ as
\begin{align}
\mathrm{E}\left[\bar{P}_{R}^{\mathrm{Group}}\right] & =G\mathrm{E}\left[\left\Vert \mathbf{h}_{RI,1}\right\Vert ^{2}\right]^{2}+G\left(G-1\right)\mathrm{E}\left[\left\Vert \mathbf{h}_{RI,1}\right\Vert \right]^{4}.
\end{align}
Making use of the moment of $\chi_{2N_{G}}$ distribution, we have
that $\mathrm{E}\left[\left\Vert \mathbf{h}_{RI,1}\right\Vert \right]=\Gamma\left(N_{G}+\frac{1}{2}\right)/\Gamma\left(N_{G}\right)$
and $\mathrm{E}\left[\left\Vert \mathbf{h}_{RI,1}\right\Vert ^{2}\right]=N_{G}$
where $\Gamma\left(x\right)$ refers to the gamma function. With $G=N_{I}/N_{G}$
and the expressions of moments, $\mathrm{E}\left[\bar{P}_{R}^{\mathrm{Group}}\right]$
is given by
\begin{align}
\mathrm{E}\left[\bar{P}_{R}^{\mathrm{Group}}\right] & =N_{I}N_{G}+\frac{N_{I}}{N_{G}}\left(\frac{N_{I}}{N_{G}}-1\right)\left(\frac{\Gamma\left(N_{G}+\frac{1}{2}\right)}{\Gamma\left(N_{G}\right)}\right)^{4}.\label{eq:scaling law Rayleigh-group}
\end{align}

\begin{figure}[t]
\begin{centering}
\includegraphics[width=8.5cm]{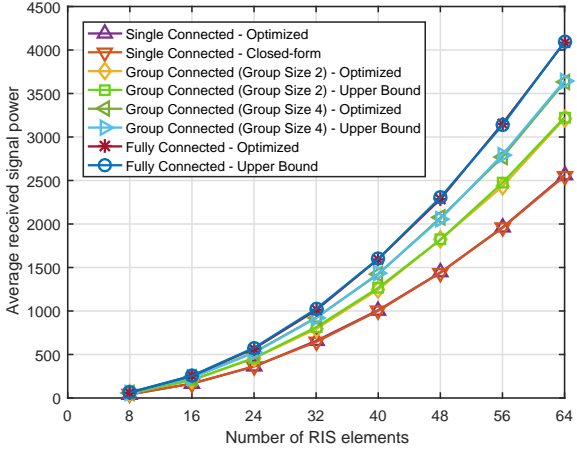}
\par\end{centering}
\caption{\label{fig:upperbound_Optimization}Average received signal power
versus the number of RIS elements.}
\end{figure}

The single and fully connected reconfigurable impedance networks can
be viewed as two special cases of the group connected reconfigurable
impedance network, i.e. the group size $N_{G}=1$ and $N_{G}=N_{I}$.
Therefore, from \eqref{eq:scaling law Rayleigh-group}, we can straightforwardly
derive that 
\begin{align}
\mathrm{E}\left[P_{R}^{\mathrm{Single}}\right] & =N_{I}+N_{I}\left(N_{I}-1\right)\Gamma\left(1.5\right)^{4},\label{eq:scaling law Rayleight-single}\\
\mathrm{E}\left[\bar{P}_{R}^{\mathrm{Fully}}\right] & =N_{I}^{2}.\label{eq:scaling law Rayleigh-fully}
\end{align}
From \eqref{eq:scaling law Rayleight-single}, we can deduce that
$\mathrm{E}\left[P_{R}^{\mathrm{Single}}\right]\rightarrow N_{I}^{2}\Gamma\left(1.5\right)^{4}=N_{I}^{2}\pi^{2}/16$
when $N_{I}\rightarrow\infty$, which is consistent with existing
results \cite{2019_TWC_IRS_QWu} indicating that the RIS has a squared
power gain. For the single, group, and fully connected reconfigurable
impedance networks, we directly optimize $\boldsymbol{\Theta}$ to
maximize the received signal power using the quasi-Newton method as
detailed in the next section. Using the Monte Carlo method, we optimize
$\boldsymbol{\Theta}$ for each channel realization and find the average
received signal power. The comparisons between the optimized result
and the upper bound (or the closed-form solution) for the single,
group, and fully connected reconfigurable impedance networks are shown
in Fig. \ref{fig:upperbound_Optimization}. We can observe that the
optimized average received signal power is the same as the upper bound
for the group and fully connected cases, which shows the upper bounds
\eqref{eq:upper bound of fully} and \eqref{eq:upper bound in PR}
are tight. Therefore, we can conclude that $\mathrm{E}\left[P_{R}^{\mathrm{Group}}\right]=\mathrm{E}\left[\bar{P}_{R}^{\mathrm{Group}}\right]$
and $\mathrm{E}\left[P_{R}^{\mathrm{Fully}}\right]=\mathrm{E}\left[\bar{P}_{R}^{\mathrm{Fully}}\right]$.

\begin{figure}[t]
\begin{centering}
\includegraphics[width=8.5cm]{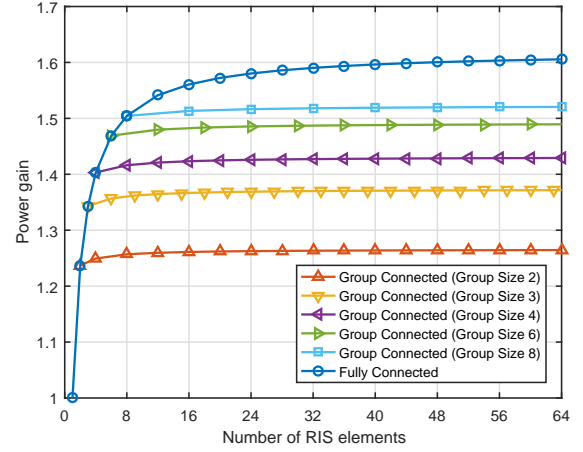}
\par\end{centering}
\caption{\label{fig:powerGain}Power gain of the group connected and fully
connected reconfigurable impedance networks over the single connected
reconfigurable impedance network.}
\end{figure}

Additionally, we can observe that 1) the group connected case achieves
higher average received signal power than the single connected case,
2) the larger the group size is, the higher the average received signal
power is, and 3) the fully connected case achieves the highest power.
The higher received power of the fully and group connected architectures
in Rayleigh fading channel comes from their ability to adjust both
the phases and the magnitudes of the impinging waves. To quantify
the increase in the received signal power, we can find the power gain
of the group and fully connected cases over the single connected case,
which are respectively given by
\begin{align}
\mathcal{G}^{\mathrm{Group}} & =\frac{N_{G}+\frac{\left(N_{I}-N_{G}\right)}{N_{G}^{2}}\left(\frac{\Gamma\left(N_{G}+\frac{1}{2}\right)}{\Gamma\left(N_{G}\right)}\right)^{4}}{1+\left(N_{I}-1\right)\Gamma\left(1.5\right)^{4}},\label{eq:gain}\\
\mathcal{G}^{\mathrm{Fully}} & =\frac{N_{I}}{1+\left(N_{I}-1\right)\Gamma\left(1.5\right)^{4}}.
\end{align}
When $N_{I}\rightarrow\infty$, the limits of the power gain are
\begin{align}
\underset{N_{I}\rightarrow\infty}{\mathrm{lim}}\mathcal{G}^{\mathrm{Group}} & =\frac{1}{N_{G}^{2}}\left(\frac{\Gamma\left(N_{G}+\frac{1}{2}\right)}{\Gamma\left(N_{G}\right)\Gamma\left(1.5\right)}\right)^{4}\label{eq:Fully Gain}\\
\underset{N_{I}\rightarrow\infty}{\mathrm{lim}}\mathcal{G}^{\mathrm{Fully}} & =\frac{1}{\Gamma\left(1.5\right)^{4}}=\frac{16}{\pi^{2}}.
\end{align}
The power gains $\mathcal{G}^{\mathrm{Group}}$ and $\mathcal{G}^{\mathrm{Fully}}$
versus $N_{I}$ are shown in Fig. \ref{fig:powerGain}. We can observe
that the power gain increases with the group size. For $N_{G}=2$,
3, 4, 6, 8, the power gain is around 1.26, 1.37, 1.43, 1.49, 1.52,
respectively. However, increasing the group size cannot increase the
power gain without limit. For the fully connected case (the maximum
group size), we can find that the limit of the power gain is around
1.62, which is consistent with \eqref{eq:Fully Gain}. Comparing $\mathcal{G}^{\mathrm{Group}}$
and $\mathcal{G}^{\mathrm{Fully}}$, we find that small group size
$N_{G}$, such as 2, 3, 4, achieves satisfactory power gain while
maintaining low complexity. Therefore, the group connected reconfigurable
impedance network with small group size is more useful in practice.

Given the same average received signal power, the number of RIS elements
required by the group connected or fully connected reconfigurable
impedance network (denoted as $N_{I}^{\mathrm{Group}}$ and $N_{I}^{\mathrm{Fully}}$)
is less than that required by the single connected reconfigurable
impedance network (denoted as $N_{I}^{\mathrm{Single}}$). When the
number of RIS elements is large, from \eqref{eq:gain}, we can deduce
that
\begin{equation}
\left(N_{I}^{\mathrm{Single}}\right)^{2}=\mathcal{G}^{\mathrm{Group}}\left(N_{I}^{\mathrm{Group}}\right)^{2}.
\end{equation}
The percentage decrease in the number of RIS elements is given by
\begin{equation}
\delta=\frac{N_{I}^{\mathrm{Single}}-N_{I}^{\mathrm{Group}}}{N_{I}^{\mathrm{Single}}}=1-\frac{1}{\sqrt{\mathcal{G}^{\mathrm{Group}}}},
\end{equation}
which can also be applied to the fully connected case by replacing
$\mathcal{G}^{\mathrm{Group}}$ with $\mathcal{G}^{\mathrm{Fully}}$.
Therefore, for $N_{G}=2$, 3, 4, 6, 8, we have $\delta=11\%$, 14\%,
16\%, 18\%, 19\%, respectively. For the fully connected case, we have
$\delta=21\%$. Such reduction in the number of RIS elements is beneficial
for reducing the cost and area of RIS, especially when the number
of RIS elements is large. Similarly, we find for small group sizes
$N_{G}$, such as 2, 3, 4, they achieve the greatest relative RIS
element reduction while maintaining low complexity, demonstrating
that the group connected reconfigurable impedance network with small
group size is more useful in practice.

To conclude, comparing the performance of the three RIS architectures
in LoS and Rayleigh fading channels, we show the gains of fully and
group connected cases appear when the channel gains (the modulus)
of $\mathbf{h}_{RI}$ and $\mathbf{h}_{IT}$ are linear independent.

\section{Performance Evaluation}

In this section, we formulate the received signal power maximization
in an RIS aided SISO system with the fully connected and group connected
reconfigurable impedance networks, and evaluate the performance in
a realistic channel model. The received signal power is given by $P_{R}=P_{T}\left\Vert h_{RT}+\mathbf{h}_{RI}\boldsymbol{\Theta}\mathbf{h}_{IT}\right\Vert ^{2}.$
We first consider the group connected reconfigurable impedance network
in the RIS, corresponding to the constraints \eqref{eq:diag(group)},
\eqref{eq:group gamma unitary constraint}. Maximizing $P_{T}\left\Vert h_{RT}+\mathbf{h}_{RI}\boldsymbol{\Theta}\mathbf{h}_{IT}\right\Vert ^{2}$
is equivalent to maximizing $\left\Vert \mathbf{h}_{RI}\boldsymbol{\Theta}\mathbf{h}_{IT}\right\Vert ^{2}$
since $\mathbf{h}_{RI}\boldsymbol{\Theta}\mathbf{h}_{IT}$ can always
be made in phase with $h_{RT}$. Therefore, we can equivalently formulate
the received signal power maximization problem with the group connected
reconfigurable impedance network as
\begin{align}
\underset{\boldsymbol{\Theta},\boldsymbol{\Theta}_{g}}{\mathsf{\mathrm{max}}}\;\; & \left\Vert \mathbf{h}_{RI}\boldsymbol{\Theta}\mathbf{h}_{IT}\right\Vert ^{2}\label{eq:MISO OP3 Objective}\\
\mathsf{\mathrm{s.t.}}\;\;\; & \boldsymbol{\Theta}=\mathrm{diag}\left(\boldsymbol{\Theta}_{1},\boldsymbol{\Theta}_{2},...,\boldsymbol{\Theta}_{G}\right),\label{eq:MISO OP3 Constraint-1}\\
 & \boldsymbol{\Theta}_{g}^{H}\boldsymbol{\Theta}_{g}=\boldsymbol{\mathrm{I}},\:\forall g,\\
 & \boldsymbol{\Theta}_{g}=\boldsymbol{\Theta}_{g}^{T},\:\forall g.\label{eq:MISO OP3 Constraint-2}
\end{align}
The constraints \eqref{eq:MISO OP3 Constraint-1}-\eqref{eq:MISO OP3 Constraint-2}
indicate that $\boldsymbol{\Theta}$ is a block diagonal matrix with
each block being a complex symmetric unitary matrix, which makes the
optimization difficult. To handle that, we leverage the relationship
between the scattering matrix $\boldsymbol{\Theta}_{g}$ and the reactance
matrix $\mathbf{X}_{I,g}$, as provided in \eqref{eq:transform S to X BLOCK},
to equivalently rewrite the problem \eqref{eq:MISO OP3 Objective}-\eqref{eq:MISO OP3 Constraint-2}
as
\begin{align}
\underset{\boldsymbol{\Theta},\mathbf{X}_{I,g}}{\mathsf{\mathrm{max}}}\;\; & \left\Vert \mathbf{h}_{RI}\boldsymbol{\Theta}\mathbf{h}_{IT}\right\Vert ^{2}\label{eq:MISO OP4 Objective}\\
\mathsf{\mathrm{s.t.}}\;\;\; & \boldsymbol{\Theta}=\mathrm{diag}\left(\boldsymbol{\Theta}_{1},\boldsymbol{\Theta}_{2},...,\boldsymbol{\Theta}_{G}\right),\label{eq:MISO OP4 Contraint-1}\\
 & \boldsymbol{\Theta}_{g}=\left(j\mathbf{X}_{I,g}+Z_{0}\mathbf{I}\right)^{-1}\left(j\mathbf{X}_{I,g}-Z_{0}\mathbf{I}\right),\:\forall g,\label{eq:MISO OP4 Contraint-12}\\
 & \mathbf{X}_{I,g}=\mathbf{X}_{I,g}^{T},\:\forall g,\label{eq:MISO OP4 Contraint-2}
\end{align}
which can be transformed to an unconstrained optimization problem.
Specifically, substituting \eqref{eq:MISO OP4 Contraint-1} and \eqref{eq:MISO OP4 Contraint-12}
into the objective \eqref{eq:MISO OP4 Objective}, we can express
the objective \eqref{eq:MISO OP4 Objective} as a function of $\mathbf{X}_{I,g}$
$\forall g$. Since $\mathbf{X}_{I,g}$ can be an arbitrary $N_{G}\times N_{G}$
real symmetric matrix, $\mathbf{X}_{I,g}$ is a function of the $N_{G}\left(N_{G}+1\right)/2$
entries in the upper triangular part, i.e. $\left[\mathbf{X}_{I,g}\right]_{i,j}$
for $i\leq j$, and there is no constraint for $\left[\mathbf{X}_{I,g}\right]_{i,j}$
for $i\leq j$. Therefore, we can express the objective \eqref{eq:MISO OP4 Objective}
as a function of $\left[\mathbf{X}_{I,g}\right]_{i,j}$ for $i\leq j$
and all $g$. Subsequently we can transform the problem \eqref{eq:MISO OP4 Objective}-\eqref{eq:MISO OP4 Contraint-2}
to an unconstrained problem which optimizes $N_{I}\left(N_{G}+1\right)/2$
unconstrained variables $\left[\mathbf{X}_{I,g}\right]_{i,j}$ for
$i\leq j$ and all $g$. To solve the unconstrained optimization problem,
we can use the quasi-Newton method in MATLAB to directly optimize
$\left[\mathbf{X}_{I,g}\right]_{i,j}$ for $i\leq j$ and all $g$,
and then find a stationary point of the problem \eqref{eq:MISO OP4 Objective}-\eqref{eq:MISO OP4 Contraint-2}.
Using the quasi-Newton method with BFGS update, the computational
complexity for each iteration is $\mathcal{O}\left(N_{I}^{2}\left(N_{G}+1\right)^{2}/4\right)$
\cite{nocedal2006numerical_optimization}. The single and fully connected
reconfigurable impedance networks can be viewed as two special cases
of the group connected reconfigurable impedance network, i.e. the
group size $N_{G}=1$ and $N_{G}=N_{I}$. Therefore, we can follow
the same approach to solve the received signal power maximization
problem with single and fully connected cases.

\begin{figure}[t]
\begin{centering}
\includegraphics[width=6.5cm]{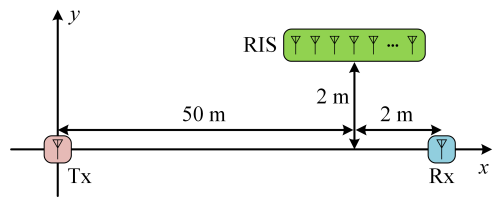}
\par\end{centering}
\caption{\label{fig:channel}2D coordinate system for the SISO RIS aided system.}
\end{figure}

We now evaluate the performance of the RIS aided SISO system with
the single, fully, and group connected reconfigurable impedance networks.
We consider a two-dimensional (2D) coordinate system as shown in Fig.
\ref{fig:channel}. A single-antenna transmitter is located at (0,
0) and a single-antenna receiver is located at (52, 0). A uniform
linear array (ULA) at the RIS are located in $x$-axis. The antenna
spacing is half wavelength and the center of the array is located
at (50, 2). The distance-dependent pathloss model is given by
\begin{equation}
L_{ij}\left(d_{ij}\right)=C_{0}\left(\frac{d_{ij}}{D_{0}}\right)^{-\alpha_{ij}}
\end{equation}
where $C_{0}$ refers to the pathloss at the reference distance $D_{0}=1$
meter (m), $d_{ij}$ refers to the distance, and $\alpha_{ij}$ refers
to the pathloss exponent for $ij\in\left\{ RT,RI,IT\right\} $. For
the small-scale fading, we assume that the transmitter-receiver and
RIS-receiver channels are both Rayleigh fading channels, and assume
that the transmitter-RIS channel is Rician fading as provided in \eqref{eq:Rician Channel Model}.
We set $\alpha_{RT}=3.5$, $\alpha_{IT}=2$, $\alpha_{RI}=2.8$, $C_{0}=-30$
dB, and $P_{T}=10$ W.

\begin{figure}[t]
\begin{centering}
\includegraphics[width=8.5cm]{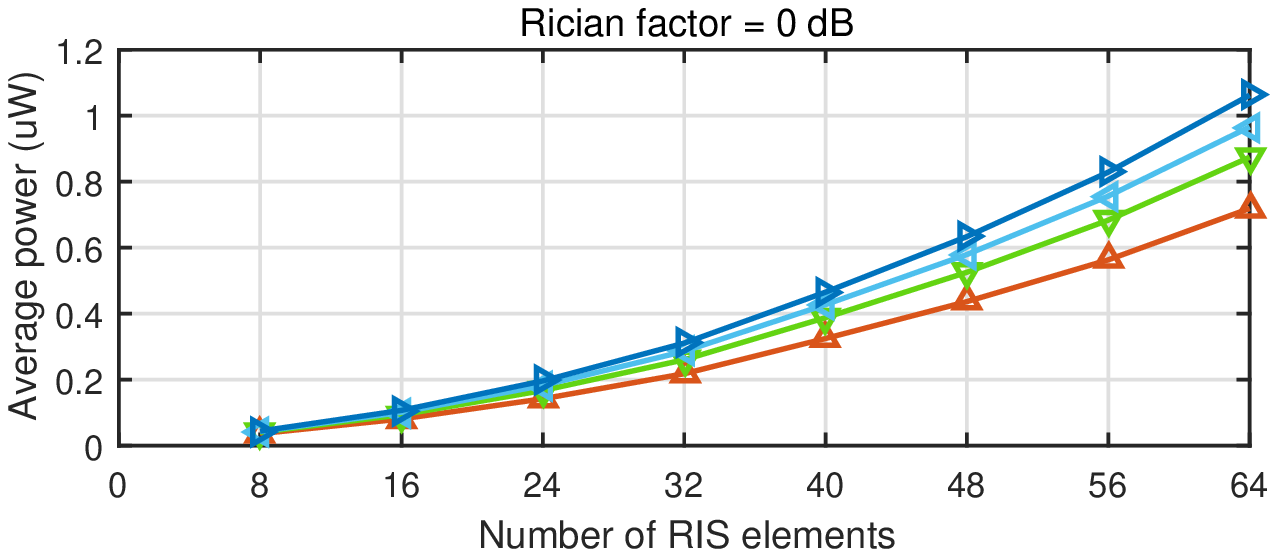}
\par\end{centering}
\begin{centering}
\includegraphics[width=8.5cm]{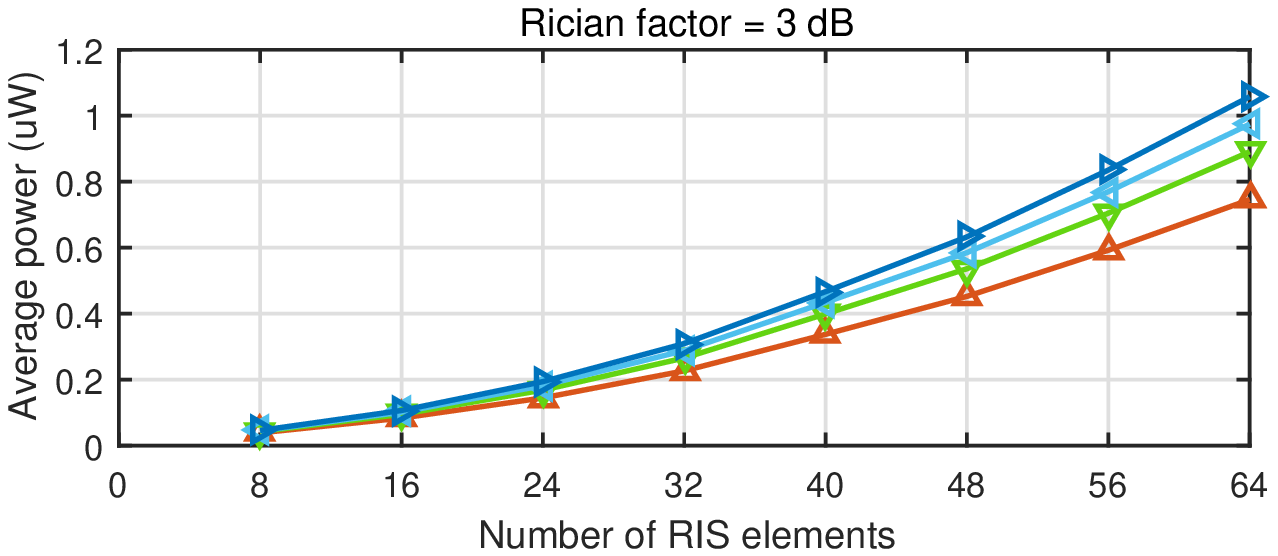}
\par\end{centering}
\begin{centering}
\includegraphics[width=8.5cm]{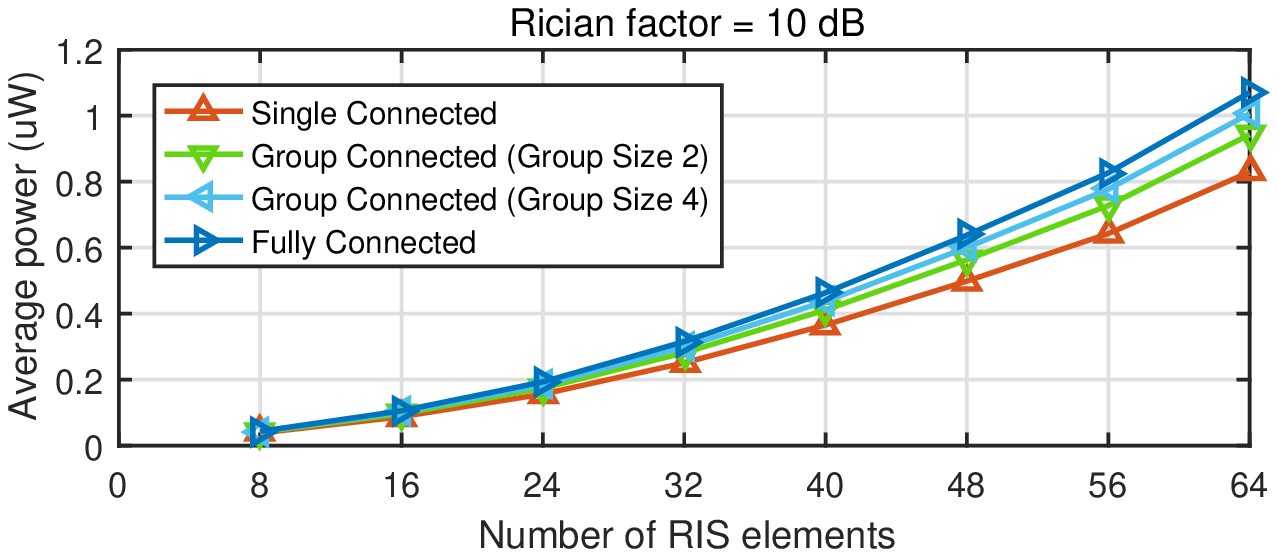}
\par\end{centering}
\caption{\label{fig:average_power_Rician}Average received signal power versus
the number of RIS elements with different Rician factors of the transmitter-RIS
channel.}
\end{figure}

Using the Monte Carlo method, we compute the average received signal
power achieved after optimizing the single, fully, and group connected
reconfigurable impedance networks. The average received signal power
versus the number of RIS elements with different Rician factors of
the transmitter-RIS channel is shown in Fig. \ref{fig:average_power_Rician}.
We can make the following observations. \textit{First}, the group
connected reconfigurable impedance network achieves a higher received
signal power than the single connected reconfigurable impedance network.
The fully connected reconfigurable impedance network achieves the
highest received signal power. This demonstrates the benefit of the
group and fully connected reconfigurable impedance networks. \textit{Second},
the larger the group size is, the higher the received signal power
is, which indicates that we can trade complexity for signal power
enhancement. \textit{Last}, the received signal power of the fully
connected reconfigurable impedance network does not change with the
Rician factor. Indeed, it can be deduced from the scaling laws \eqref{eq:scaling law-LOS}
and \eqref{eq:scaling law Rayleigh-fully} in that the received signal
power is always $N_{I}^{2}$. However, the received signal power of
the single connected case increases with the Rician factor. Indeed,
we can deduce from the scaling laws \eqref{eq:scaling law-LOS} and
\eqref{eq:scaling law Rayleight-single} in that the single connected
case achieves a higher power in the LoS channel compared to Rayleigh
fading channels.

\begin{figure}[t]
\begin{centering}
\includegraphics[width=8.5cm]{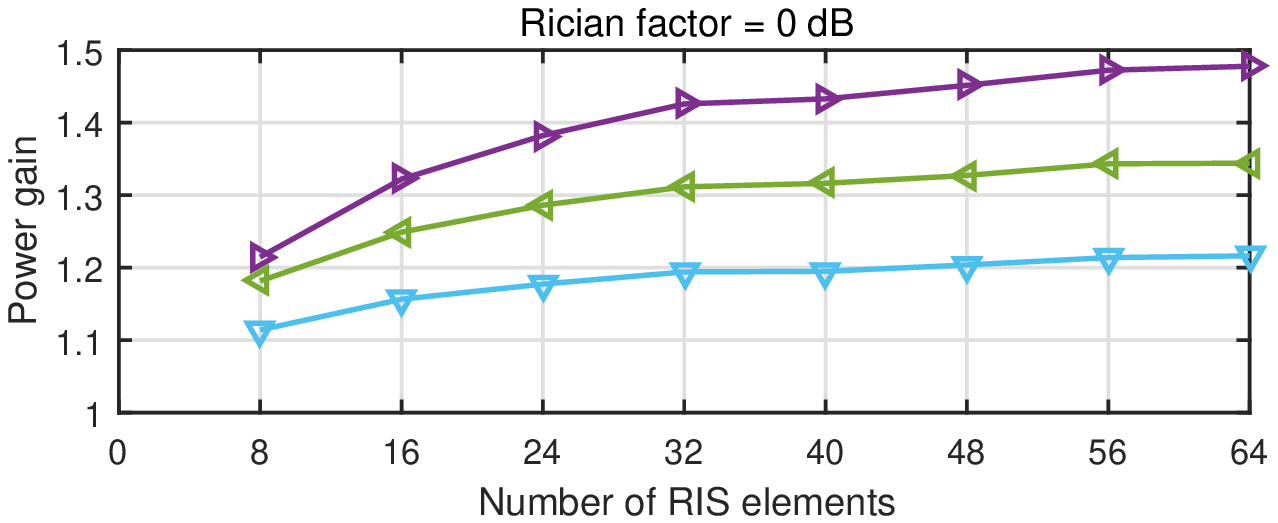}
\par\end{centering}
\begin{centering}
\includegraphics[width=8.5cm]{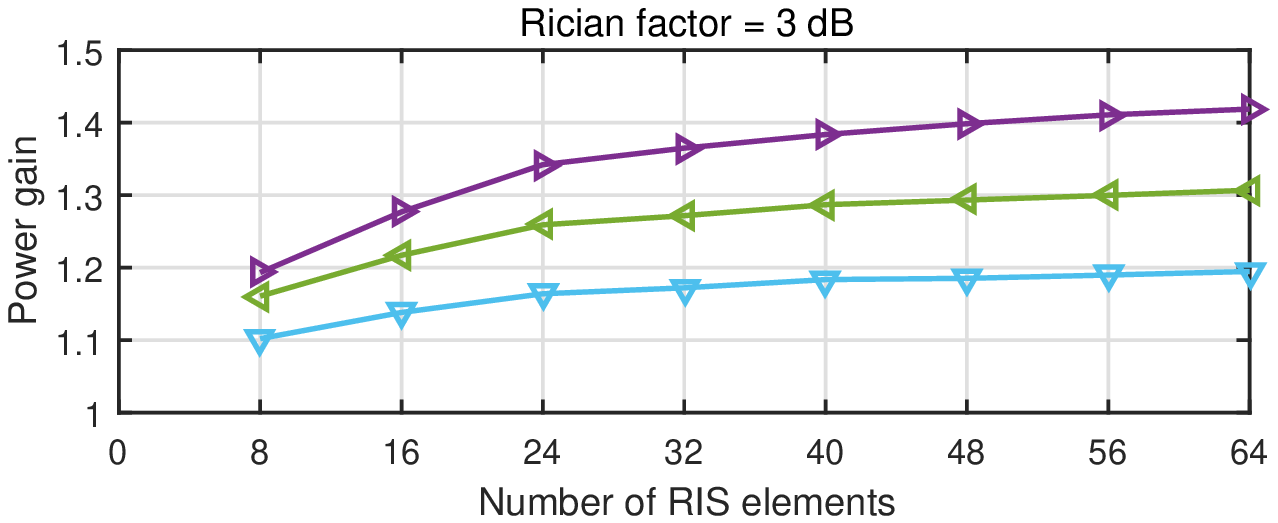}
\par\end{centering}
\begin{centering}
\includegraphics[width=8.5cm]{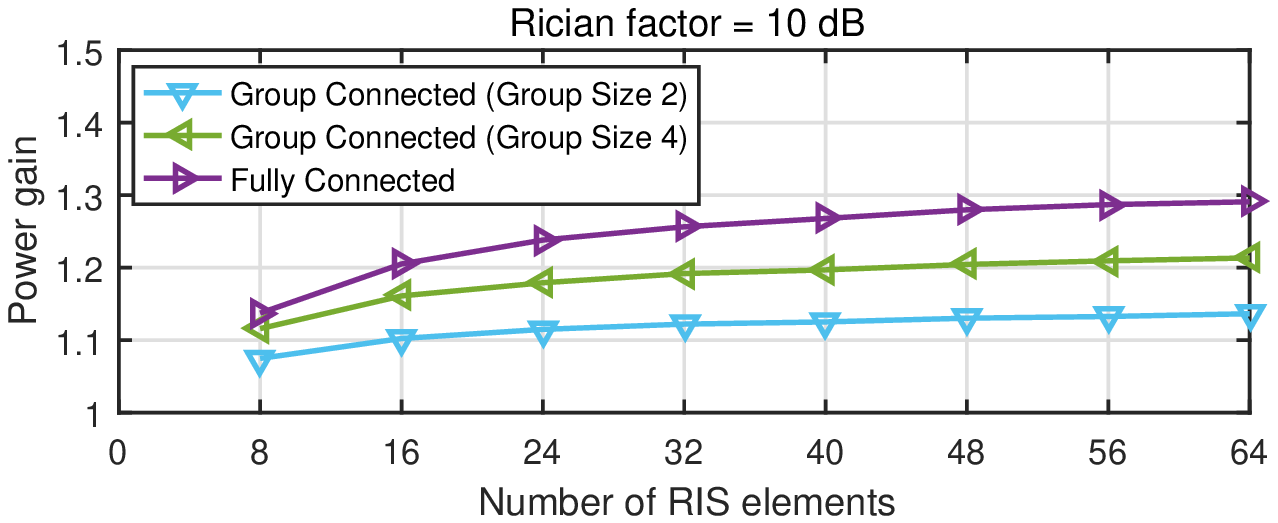}
\par\end{centering}
\caption{\label{fig:Power_gain_Rician}Power gains of the group connected and
fully connected reconfigurable impedance networks over the single
connected reconfigurable impedance network with different Rician factors
of the transmitter-RIS channel.}
\end{figure}

We also plot the power gains of the group and fully connected reconfigurable
impedance networks over the single connected reconfigurable impedance
network with different Rician factors of the transmitter-RIS channel
in Fig. \ref{fig:Power_gain_Rician}. We find that the power gain
decreases with the Rician factor. To show the tradeoff between performance
and complexity for the single, group, and fully connected reconfigurable
impedance networks, we quantify the complexity from two perspectives,
1) the circuit topology complexity, which refers to the number of
reconfigurable impedance components in the reconfigurable impedance
network, and 2) the optimization computational complexity, which refers
to the computational complexity for optimizing the reconfigurable
impedance networks with different constraints of $\boldsymbol{\Theta}$.
A comprehensive comparison of the single, group, and fully connected
reconfigurable impedance networks in terms of the power gain at different
Rician factors, the circuit topology complexity, and the optimization
computational complexity is summarized in Table. \ref{tab:Comprehensive-Comparison-of}.
We can conclude that the power gain can be enhanced by introducing
more reconfigurable components and more optimization computations,
i.e. trading circuit topology complexity and optimization computational
complexity for performance enhancement. Particularly, the group connected
reconfigurable impedance network with group size of 2 achieves a good
tradeoff between complexity and performance enhancement, which uses
half more reconfigurable impedance components and 1.25 times more
computations to improve the received signal power by around 20\%.
In addition, given the same received signal power, using the group
connected reconfigurable impedance network can reduce the number of
RIS elements. For example, for the group size of 2 and 4 with Rician
factor of 0 dB, the number of RIS elements can be reduced by 9.5\%
and 13.6\%, respectively. Such reduction of the number of RIS elements
is beneficial for reducing the cost and area of RIS, especially when
the number of RIS elements is large.

\begin{table*}[t]
\caption{\label{tab:Comprehensive-Comparison-of}Comprehensive Comparison of
single connected, group connected, and fully connected reconfigurable
impedance network.}

\begin{centering}
\begin{tabular}{|c|c|c|c|c|c|c|}
\hline 
\multirow{2}{*}{} &  & \multicolumn{3}{c|}{{\scriptsize{}Power Gain}} & {\scriptsize{}Circuit } & {\scriptsize{}Optimization }\tabularnewline
\cline{3-5} \cline{4-5} \cline{5-5} 
 & {\scriptsize{}Group Size} & {\scriptsize{}Rician } & {\scriptsize{}Rician } & {\scriptsize{}Rician } & {\scriptsize{}Topology } & {\scriptsize{}Computational}\tabularnewline
 &  & {\scriptsize{}Factor 0 dB} & {\scriptsize{}Factor 3 dB} & {\scriptsize{}Factor 10 dB} & {\scriptsize{}Complexity} & {\scriptsize{}Complexity}{\footnotesize{}$^{\dagger}$}\tabularnewline
\hline 
{\scriptsize{}Single Connected} & {\scriptsize{}1} & {\scriptsize{}1} & {\scriptsize{}1} & {\scriptsize{}1} & {\scriptsize{}$N_{I}$} & {\scriptsize{}$\mathcal{O}\left(N_{I}^{2}\right)$}\tabularnewline
\hline 
{\scriptsize{}Group Connected} & {\scriptsize{}2} & {\scriptsize{}1.22} & {\scriptsize{}1.20} & {\scriptsize{}1.14} & {\scriptsize{}1.5$N_{I}$} & {\scriptsize{}$\mathcal{O}\left(2.25N_{I}^{2}\right)$}\tabularnewline
\hline 
{\scriptsize{}Group Connected} & {\scriptsize{}4} & {\scriptsize{}1.34} & {\scriptsize{}1.31} & {\scriptsize{}1.21} & {\scriptsize{}2.5$N_{I}$} & {\scriptsize{}$\mathcal{O}\left(6.25N_{I}^{2}\right)$}\tabularnewline
\hline 
{\scriptsize{}Fully Connected} & {\scriptsize{}$N_{I}$} & {\scriptsize{}1.48} & {\scriptsize{}1.42} & {\scriptsize{}1.30} & {\scriptsize{}$N_{I}\left(N_{I}+1\right)/2$} & {\scriptsize{}$\mathcal{O}\left(N_{I}^{2}\left(N_{I}+1\right)^{2}/4\right)$}\tabularnewline
\hline 
\end{tabular}
\par\end{centering}
{\scriptsize{}$^{\dagger}$ The computational complexity for each
iteration using quasi-Newton method with BFGS update.}{\scriptsize\par}
\end{table*}

\section{Conclusions and Future Work}

We use scattering parameter network analysis to derive a physical
and EM compliant yet straightforward and tractable RIS aided communication
model. The proposed general RIS aided communication model fully considers
the effects of impedance mismatching and mutual coupling at the transmitter,
RIS, and receiver, and thus is more comprehensive than the conventional
RIS aided communication model \cite{2020_CM_TowardIRS_QWu}, \cite{rajatheva2020white},
\cite{2019_TWC_RIS_CHuang}-\cite{2020_WCL_IRS_DRL} which does not
consider these effects. Furthermore, the proposed general model can
be reduced to the conventional RIS aided communication model \cite{2020_CM_TowardIRS_QWu},
\cite{rajatheva2020white}, \cite{2019_TWC_RIS_CHuang}-\cite{2020_WCL_IRS_DRL}
under special conditions.

Using the proposed RIS model we also develop new RIS architectures
based on group and fully connected reconfigurable impedance networks,
which are more general and efficient than previously utilized single
connected architecture \cite{2020_CM_TowardIRS_QWu}, \cite{rajatheva2020white},
\cite{2019_TWC_RIS_CHuang}-\cite{2020_WCL_IRS_DRL}. In sharp contrast
with the single connected architecture that only adjusts the phases
of the impinging waves using a diagonal scattering matrix, our proposed
group and fully connected architectures enable scattering matrices
to be block diagonal or full and can adjust not only the phases but
also the magnitudes of the impinging waves, so as to provide better
performance in RIS aided systems.

We derive the scaling law of the received signal power of a SISO RIS
aided system as a function of the number of RIS elements in both LoS
and Rayleigh fading channels. It shows that using fully and group
connected reconfigurable impedance networks can increase the received
signal power by up to 62\% compared with the single connected case.
It also indicates that given the same received signal power, using
fully connected and group connected reconfigurable impedance networks
can reduce the number of RIS elements by up to 21\%. We also formulate
the received signal power maximization problem in the SISO RIS aided
system and evaluate the received signal power in a realistic model
with distance-dependent pathloss and Rician fading channel. The numerical
results show that the fully and group connected reconfigurable impedance
networks can increase the received signal power by up to 48\% and
34\%, respectively.

Future research avenues include, but are not limited to, the following
areas:

1) Developing efficient channel estimation methods. For the proposed
RIS aided communication model with perfect matching and no mutual
coupling, the channel matrix \eqref{eq:perfectmatchign H} is exactly
the same as the conventional RIS aided communication model \cite{2020_CM_TowardIRS_QWu},
\cite{rajatheva2020white}, \cite{2019_TWC_RIS_CHuang}-\cite{2020_WCL_IRS_DRL},
so that we can use the channel estimation methods for conventional
RIS aided communication model \cite{2019_TWC_IRS_QWu}, \cite{2019_ICASSP_IRS_WET}
to estimate the channel matrix. For the general RIS aided communication
model, from the channel matrix expression \eqref{eq:H general}, we
need to first measure the impedance mismatch $\mathbf{\Gamma}_{T}$
and $\mathbf{\Gamma}_{R}$ and mutual coupling $\mathbf{S}_{TT}$,
$\mathbf{S}_{II}$, and $\mathbf{S}_{RR}$ by a vector network analyzer
and then use the channel estimation methods \cite{2019_TWC_IRS_QWu},
\cite{2019_ICASSP_IRS_WET} to estimate the channel matrix. In the
future, we can develop more efficient channel estimation methods for
the proposed RIS aided communication models.

2) Extending to multi-user and multi-cell scenarios. Previous work
on multi-user \cite{2019_TWC_IRS_QWu} and multi-cell scenarios \cite{2020_TWC_IRS_MIMO_Multicell}
only use RIS with the single connected architecture to minimize the
transmit power and maximize the weight sum rate, respectively. In
the future, we can consider using the fully and group connected architectures,
which are more general than the single connected architecture, to
further decrease the transmit power in multi-user scenario and increase
the weighted sum rate in multi-cell scenario.

3) Optimizing with discrete values of $\boldsymbol{\Theta}$. For
the single connected architecture, we have $\boldsymbol{\Theta}=\mathrm{diag}\left(e^{j\theta_{1}},e^{j\theta_{2}},...,e^{j\theta_{N_{I}}}\right)$
and we can restrict the continuous $\theta_{n_{I}}\in\left[0,2\pi\right]$
to discrete values, which is called discrete phase shifts. The optimization
with discrete phase shifts has been well studied in \cite{2019_PACRIM_IRS_DiscretePhaseShift},
\cite{2020_ToC_IRS_Discrete_Phase}. Inspired by the discrete phase
shifts, in the future we can consider the design and optimization
of discrete values of matrix $\boldsymbol{\Theta}$ for the fully
and group connected architectures.

\section*{Appendix}

Scattering parameter network theory \cite{pozar2009microwave} is
useful to model and analyze wireless systems. This theory has been
used to accurately characterize MIMO wireless systems in previous
work \cite{2004_TWC_Jensen}, \cite{2004_TAP_Jensen}, \cite{2006_TAP_BKLau_impedance_matching}
for example. In this appendix, we briefly review the concept of reflection
coefficient and scattering parameters to help provide some background
\cite{pozar2009microwave}.

\subsection*{A. Reflection Coefficient}

\begin{figure}[t]
\begin{centering}
\includegraphics[width=5cm]{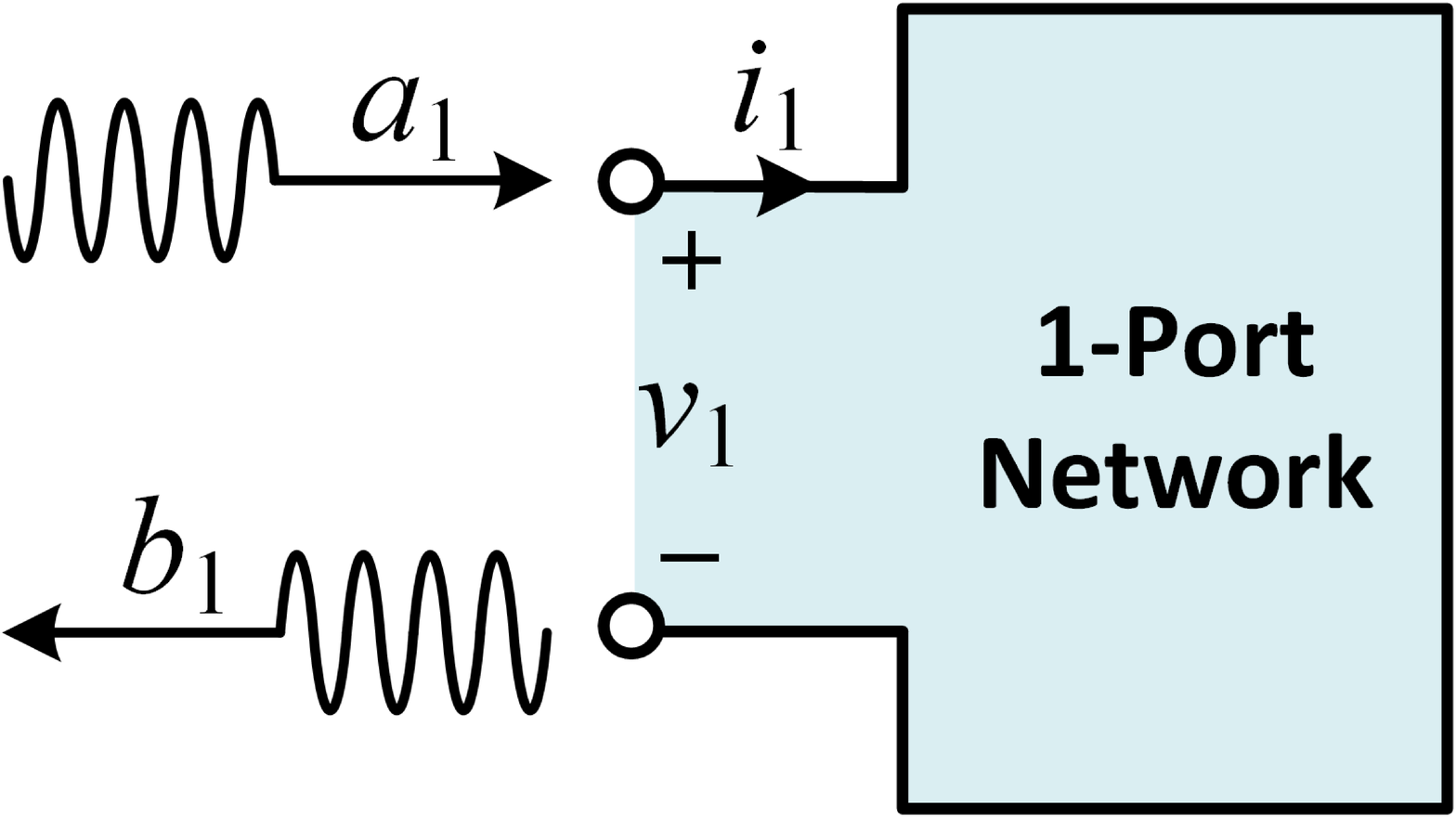}
\par\end{centering}
\caption{\label{fig:A-transmission-line}An arbitrary $1$-port network.}
\end{figure}

Consider an arbitrary 1-port network as shown in Fig. \ref{fig:A-transmission-line}.
The 1-port network can be a source impedance, or a load impedance,
or an antenna impedance, and it can be constructed from wires, transmission
lines, circuits, antennas, or more generally it can be any linear
electromagnetic system \cite{pozar2009microwave}. Assume that an
incident voltage wave, denoted as $a_{1}$, is input into the 1-port
network. The incident voltage wave will be reflected by the 1-port
network and subsequently a reflected voltage wave, denoted as $b_{1}$,
is generated. Therefore, the voltage across the port, denoted as $v_{1}$,
is the sum of the incident and reflected voltage waves, i.e. $v_{1}=a_{1}+b_{1}$,
and the current through the port, denoted as $i_{1}$, is described
by $i_{1}=\left(a_{1}-b_{1}\right)/Z_{0}$ where $Z_{0}$ is a chosen
reference impedance and usually it is set as $Z_{0}=50\:\Omega$.
We define the ratio of the reflected and incident voltage waves as
the reflection coefficient of the 1-port network, which is denoted
as $\Gamma$ and given by
\begin{equation}
\Gamma=\frac{b_{1}}{a_{1}}=\frac{Z-Z_{0}}{Z+Z_{0}},
\end{equation}
where $Z=v_{1}/i_{1}$ denotes the input impedance of the 1-port network.
$\Gamma$ and $Z$ have a one-to-one correspondence relationship so
that $\Gamma$ can completely characterize any input impedance of
a 1-port network. For a passive input impedance $\mathfrak{\Re}\left\{ Z\right\} \geq0$,
we have that $\left|\Gamma\right|\leq1$. Particularly, for a pure
reactive input impedance $\mathfrak{\Re}\left\{ Z\right\} =0$, we
have that $\Gamma=e^{j\theta}$ and $\left|\Gamma\right|=1$, which
is helpful to increase the power of scattered wave and is the key
property of RIS (the phase shift and the unit modulus constraint).

It should also be noted that the reflection coefficient can characterize
any 1-port network no matter how it is constructed. Particularly,
the reflection coefficient of an antenna characterizes how much of
the incident wave the antenna can radiate, which is an important parameter
in antenna design for wireless systems.

\begin{figure}[t]
\begin{centering}
\includegraphics[width=6.5cm]{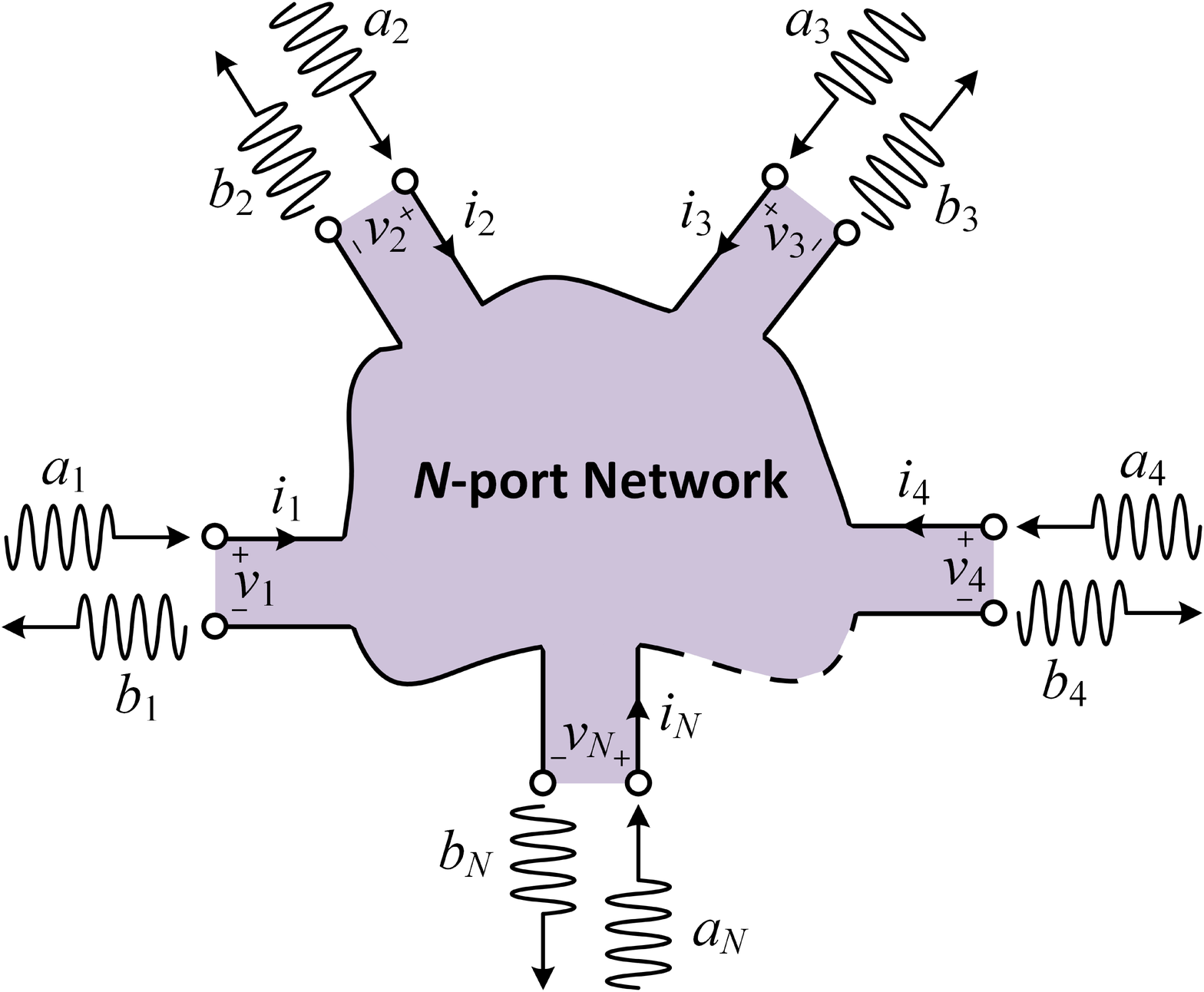}
\par\end{centering}
\caption{\label{fig:Smatrix}An arbitrary $N$-port network.}
\end{figure}

\subsection*{B. Scattering Parameters}

Generalizing the concept of the 1-port network, we consider an arbitrary
$N$-port network as shown in Fig. \ref{fig:Smatrix}, where $a_{n}$
refers to the wave incident on the $n$th port and $b_{n}$ refers
to the wave reflected from the $n$th port. Denote $\mathbf{a}=\left[a_{1},a_{2},\ldots,a_{N}\right]^{T}$
and $\mathbf{b}=\left[b_{1},b_{2},\ldots,b_{n}\right]^{T}$. The scattering
parameter matrix $\mathbf{S}$ is defined in relation to these incident
and reflected waves as $\mathbf{b}=\mathbf{S}\mathbf{a}$. Similar
to the impedance or admittance parameter matrix for an $N$-port network,
the scattering parameter matrix can completely characterize the network
as seen at its $N$ ports. While the impedance and admittance parameter
matrices relate the total voltages and currents at the ports, the
scattering parameter matrix relates the waves incident on the ports
to those reflected from the ports. In particular, at the $n$th port,
the voltage and current are related with the incident and reflected
waves by $v_{n}=a_{n}+b_{n}$ and $i_{n}=\left(a_{n}-b_{n}\right)/Z_{0}$.
Therefore, the scattering matrix $\mathbf{S}$ can be one-to-one converted
to the impedance matrix $\mathbf{Z}$ through
\begin{equation}
\mathbf{S}=\left(\mathbf{Z}+Z_{0}\mathbf{I}\right)^{-1}\left(\mathbf{Z}-Z_{0}\mathbf{I}\right).\label{eq:ZtoS}
\end{equation}
For a 1-port network, the scattering matrix $\mathbf{S}$ reduces
to a scalar, which is essentially the reflection coefficient as introduced
in Appendix A.

It should be noted that scattering parameters can characterize any
$N$-port network. It does not matter if the $N$-port network is
constructed from wires, transmission lines, circuits, antennas, or
more generally it can be any linear electromagnetic system. Particularly,
for SISO wireless systems, the single transmit antenna and single
receive antenna embedded in the wireless channel can be viewed as
a 2-port network, where the $\left[\mathbf{S}\right]_{2,1}$ parameter
in the $2\times2$ scattering matrix $\mathbf{S}$ is the channel
gain between two antennas. Moreover, the scattering parameter network
analysis can accurately characterize MIMO wireless systems, as shown
in \cite{2004_TWC_Jensen}, \cite{2004_TAP_Jensen}, \cite{2006_TAP_BKLau_impedance_matching}.
In practice, scattering parameters can be measured by a vector network
analyzer \cite{pozar2009microwave} and have been widely used in the
measurements of microwave circuit and component, antennas, and wireless
systems. To conclude, reflection coefficient and scattering parameters
are suitable and accurate for modeling wireless systems.


\end{document}